\documentstyle[12pt,preprint,psfig]{aastex}
\newcommand{\etal}{et al.\  }

\lefthead{Mart\'\i nez-Delgado \etal}
\righthead{Sagittarius}

\begin{document}

\title{Tracing out the Northern Tidal Stream of the Sagittarius Dwarf Spheroidal Galaxy}

\author{David Mart\'\i nez-Delgado} 
\affil{Instituto de Astrof\'\i sica de Canarias,
E-38205 La Laguna, Tenerife, Canary Islands, Spain}
\affil{Max-Planck-Institut fur Astronomie,
Konigstuhl,17, D-69117 Heidelberg, Germany}

\author{M. \'Angeles G\'omez-Flechoso\footnote{Present address: Universidad Europea de Madrid, E-28670
Villaviciosa de Od\'on, Madrid, Spain}} 
\affil{Geneva Observatory, Ch. des Maillettes 51, CH-1290 Sauverny,
Switzerland}

\author{Antonio Aparicio} 
\affil{Instituto de Astrof\'\i sica de Canarias,
E-38205 La Laguna, Tenerife, Canary Islands, Spain}
\affil{Departamento de Astrof\'\i sica, Universidad de La Laguna,
E38200 - La Laguna, Tenerife, Canary Islands, Spain}

\author{Ricardo Carrera}
\affil{Instituto de Astrof\'\i sica de Canarias, E-38205 La Laguna, Tenerife,
Canary Islands, Spain}

\begin{abstract}

The main aim of this paper is to report two new detections of tidal debris in the northern 
stream of the Sagittarius dwarf galaxy located at  45$\arcdeg$ and 55$\arcdeg$ 
from the center of galaxy. Our observational approach is based on deep color--magnitude diagrams, that provides accurate distances, surface brightness  and the properties of stellar population of the studied region of this tidal stream. The derived distances for these tidal debris 
wraps are 45$\pm$ 2 kpc and 54 $\pm$ 2 kpc respectively.
These detections are also strong 
observational evidence that the tidal stream discovered by the Sloan Digitized Sky Survey is tidally stripped material from the Sagittarius dwarf and support the idea  that the tidal stream completely enwraps the Milky Way in an almost polar orbit. 

We also confirm these detections running numerical simulations of the Sagittarius dwarf plus 
the Milky Way. This model reproduces the present position and 
velocity of the Sagittarius main body and presents a long tidal stream formed 
by tidal interaction with the Milky Way potential. The tidal streams of the model traces the last orbit of Sagittarius and confirms our observational detections. This model is also in good agreement with the available observations of the Sagittarius tidal stream. The comparison of our model with the positions and distances of two non-identified halo overdensities discovered by the Sloan Digitized Sky Survey and the QUEST survey shows that they are actually associated to the trailing arm of the Sagittarius tidal stream. In addition, we identify the proper motion group discovered by Arnold \& Gilmore (1992) as a piece of the Sagittarius northern stream.

We also present a method for estimating the shape of the Milky Way halo potential using numerical simulations.
From our simulations we obtain an oblateness of the Milky Way dark halo potential of 0.85, using the current database of distances and radial velocities
of the Sagittarius tidal stream. 

The color--magnitude diagram of the apocenter of Sagittarius shows that this region of the stream shares the complex star formation history observed in the main body of the galaxy. We present the first evidence for a gradient in the stellar population along the stream, possibly correlated with its different pericenter passages.

\end{abstract}


\section{INTRODUCTION}

The Sagittarius (Sgr) dwarf galaxy (Ibata, Gilmore, \& Irwin 1994), a Milky Way
 satellite in an advanced state of tidal disruption, provides a ``living'' test  
 for tidal interaction models and for galaxy formation theories. Dynamical models 
 predict that the stream associated with this galaxy should envelop the
whole Milky Way in an almost polar orbit (G\'omez-Flechoso, Fux \& Martinet 1999; 
Johnston et al. 1999; Helmi \& White 2001). As the biggest tidal stream known in the 
halo, it  can provide a powerful probe of the Milky Way dark matter halo, since the 
stream stars can be used as natural test-particles along the Sgr orbit through the 
outer regions of the halo.

Different studies have revealed stellar debris associated with Sgr located at large 
distances from its center. Mateo, Olszewski \& Morrison (1998)
have identified main sequence (MS) stars of Sgr up to 34$\arcdeg$ from its center 
from a photometric survey along its  southeast major axis. Star
counts parallel to the semiminor axis in these outer fields showed that Sgr remains 
quite narrow in this direction, suggesting that its outer portion resembles a tidal stream
that is structurally different from the main body of the galaxy (the Sgr
southern stream). Theoretical models predicted the existence of a symmetric extension 
of the Sgr tidal tail along its northwest major axis too (the Sgr northern stream: 
G\'omez-Flechoso et al. 1999; Johnston et al. 1999). However, because of this 
northern extension would cross the Galactic plane,  the first attempts to detect it 
were unsuccessful (Mateo et al. 1998; Majewski et al. 1999). The conclusion was that, 
if the northern stream did exist, it would be  completely hidden behind the plane of the Milky Way.

The first evidence of the Sgr northern stream came from the first year commissioning 
data of  the Sloan Digital Sky Survey (SDSS; Yanny \etal 2000). The SDSS team  
discovered two extended overdensities of blue A-type stars in the halo, situated 
at $\simeq 45$ kpc from the Sun (in the following, we will refer to these sub-structures 
as the SDSS stream). In addition, Ivezic et al. (2000) and Vivas et al. (2001) 
reported a large clump of RR Lyrae candidates at 
a similar distance in the same direction. These results suggested that these stellar 
populations could be associated with the remnants of a disrupted system, such as a 
dwarf galaxy. Mart\'\i nez-Delgado et al. (2001a) concluded
that these stellar populations are indeed associated with a differentiated, very low 
density stellar system situated at $50 \pm 10$ kpc from the first color--magnitude 
diagram (CMD) of a region in the SDSS stream. Dohm-Palmer et al. (2001) also 
reported an excess of red giants at this distance, as well as evidence for additional 
structures at different distances, possibly associated with multiple wraps of the 
Sgr tidal stream. Newberg et al. (2002) obtained a CMD of a 110 square degree slice 
of the SDSS stream. From its resemblance to the CMD of the Sgr main body, they 
conclude that this structure is part of this galaxy. Comparison with theoretical models of the tidal destruction of this galaxy (Ibata 
\etal 2001a; Mart\'\i nez-Delgado et al. 2001a; Helmi \& White 2001) indicates that 
the  position and distance of the SDSS stream match those expected for the Sgr northern stream. More recently, Majewski et al. (2003) present a first all-sky map of the Sgr tidal stream, traced by its M star population detected in
the Two Micron All-Sky Survey (2MASS). 

 In this paper, we report two new detections of the tidal debris of the Sgr northern stream between the region sampled by the SDSS and the main body of the galaxy.  These detections were firstly reported in Mart\'\i nez-Delgado, G\'omez-Flechoso \& Aparicio (2002a), and  provided the first observational evidence  that the SDSS stream is tidally stripped material from Sgr and that the tidal stream of this galaxy completely enwraps the Milky Way in an almost polar orbit. 

 This paper is organized as follows: in Sec. 2, the observations and the 
data reduction process are described. The new detections of Sgr northern stream
are given in Sec. 3. In Sec. 4, the numerical model and the 
simulations are described; the comparison between the model and the 
observations is also given. In Sec. 5, we explain how the observations 
of the tidal streams and
the theoretical models allow us to constrain the dark matter halo of the
Milky Way. The description of the stellar population of the Sgr stream 
is given in Sec. 6. Finally, the main conclusions are summarized 
in Sec. 7.

\section{OBSERVATIONS}

\subsection{OBSERVATIONS AND DATA REDUCTION}

The observations of the Sgr northern stream were carried out in Johnson $B$
 and Sloan $R$ filters with the 2.5 m Isaac Newton Telescope (INT) at Roque
de los Muchachos Observatory on La Palma. At the INT prime focus,
we used the Wide Field Camera (WFC), which houses
 four $2048 \times 4096$ pixel EEV
chips. The plate-scale is $0.33''$/pix, which provides a total field of about $35'
\times 35'$. Table 1 provides a
summary of the observations. 

The images were processed in the usual way. Bias and flat-field corrections
were made with IRAF. DAOPHOT and ALLFRAME (Stetson 1994) were used to obtain
the photometry of the resolved stars. Transformation into the standard
photometric system requires several observations of standard star fields in
such a way that standards are measured in all four chips of the WFC. A large number 
of standard stars were measured during the observing run to calculate both the  atmospheric
extinctions for each night and the general transformation into the standard Johnson--Cousins
system for each chip. In total, 320 measurements of 55 standards were made in total during the
observing run of 2001 June.

Transformations from instrumental magnitudes measured at the top of
the atmosphere into Johnson-Cousins magnitudes are given, for each chip, by:

\begin{equation}
(B-b_1)=24.863+0.064(B-R); ~~~~\sigma=0.010,
\end{equation}
\begin{equation}
(R-r_1)=24.628-0.105(B-R); ~~~~\sigma=0.010,
\end{equation}

\begin{equation}
(B-b_2)=25.035+0.024(B-R); ~~~~\sigma=0.019,
\end{equation}
\begin{equation}
(R-r_2)=24.663-0.049(B-R); ~~~~\sigma=0.025,
\end{equation}

\begin{equation}
(B-b_3)=25.098+0.013(B-R); ~~~~\sigma=0.022,
\end{equation}
\begin{equation}
(R-r_3)=24.826-0.066(B-R); ~~~~\sigma=0.017,
\end{equation}

\begin{equation}
(B-b_4)=24.992+0.076(B-R); ~~~~\sigma=0.010,
\end{equation}
\begin{equation}
(R-r_4)=24.863-0.097(B-R); ~~~~\sigma=0.011,
\end{equation}

\noindent where subindices refer to chips, lower-case letters stand for
instrumental magnitudes and capital letters for Johnson--Cousins magnitudes.

 The $\sigma$ values are the estimated
errors of the fits, obtained as $\sigma$=$\sigma_{*}/\sqrt{N-2}$, where $\sigma_{*}$ 
is the main square root of the data dispersion around the fit and
$N$ is the number of data points used in the fit.

Finally, estimated errors in the extinctions for each night are about
$\sigma=0.009$ and $\sigma=0.006$ for $B$ and $R$ filters respectively. Aperture 
corrections were obtained using a set
of isolated bright stars in each frame, the estimated errors being of the order of 
$\sigma=0.01$. Putting all the errors together, the total zero-point error in our 
photometry can be estimated to be about $\sigma=0.016$ for all bands.

For the final photometric list, we selected stars with $\sigma < 0.20$, $-1 < SHARP < 1$ and $0<
CHI < 2$, as provided by ALLSTAR. These criteria reject extended
objects, so the background contamination is expected to be limited to stellar-shaped objects.

In addition,  $B$ and $R$ photometry of a field situated in the Sgr main body is 
used in this study. These data are part of a project devoted to the study of the 
stellar population gradients in Sgr and were obtained in September 1998 at the du 
Pont 100$''$ telescope at Las Campanas Observatory. The data reduction and photometry 
of this project will be described in detail in a forthcoming paper. The resulting 
CMD of this field was decontaminated from Milky Way foreground stars by means of 
the technique described in Gallart, Aparicio, \& V\'\i lchez (1996), using a control 
field taken during the same observing run. This decontaminated CMD has been used to 
estimate the surface brightness of the  Sgr tidal debris reported in Sec. 3 and for 
comparison purposes throughout the paper.

\subsection{METHODOLOGY AND SEARCHING STRATEGY}

Our method is based in the analysis of wide field, deep CMDs reaching the 
old MS turn-off of the Sgr stellar population. Because of the low surface 
brightness of the Sgr northern stream ($\Sigma_{V}\sim$ 31 mag arcsec$^{-2}$; 
Mart\'\i nez-Delgado et al. 2001a), it is essential to cover a large area on 
the sky to detect the galaxy's MS turn-off as a density enhancement above the 
Galactic foreground population in the CMD. This method has already been 
successful in detecting the tidal tail of Ursa Minor dSph (Mart\'\i nez-Delgado 
et al. 2001b), the potential apocenter of the Sgr stream (Mart\'\i nez-Delgado 
et al. 2001a) and the remnants of the Sgr southern stream around the globular 
cluster Pal 12 (Mart\'\i nez-Delgado et al. 2002b). In addition, this observational approach provides also accurate distances and surface brightness in the part of tidal stream detected, providing valuable constrains to the theoretical models (see Sec. 4).

The search method is illustrated in Fig. 1, which shows three model CMDs for 
a tidal stream of a dwarf galaxy assumed to be at a heliocentric distance of 45 
kpc, overplotted on a control field CMD taken with the WFC during the observing 
run (see Sec. 3.2). The latter samples the Milky Way halo population at  galactic 
latitude $b=50\arcdeg$. The decontaminated CMD of the Sgr main body obtained in 
Sec. 2.1 was used to represent the stream stellar population and shows how it 
would be seen in the CMD for three possible surface brightness values, 
obtained by plotting different numbers of stream stars. It shows that, for the 
typical stream surface brightness, the  MS turn-off is almost the only detectable 
signature of the galaxy and is thus an unambiguous indicator of the presence of Sgr 
remnants behind the 
Milky Way foreground stars (which swamp the giant branch) and faint background 
galaxies (which dominate the star counts for ``$V$''$>23.5$ and dilute the lower 
MS). In addition,  we conclude that for a distance of 45 kpc the Sgr stream would 
be undetectable with this technique if its surface brightness were below 
$\Sigma_{V}\sim$ 32 mag arcsec$^{-2}$.

Since the Sgr northern stream position is poorly defined in the surveyed area, 
a very important aspect of this paper is to alter the search strategy pattern during 
the observations. This means doing ``real-time'' photometry as soon as the data
are obtained (see Mateo et al. 1998). This procedure was performed using 
DAOPHOT/ALLSTAR software in an Ultra workstation at the telescope. This provided a 
preliminary CMD of the target field, deep enough to decide whether Sgr tidal debris 
is present. In addition, our dynamical model of the Sgr stream (Mart\'\i nez-Delgado 
et al. 2001a) was initially assumed to select the best locations to start looking 
for the stream. This model not only fitted the projected positions of the reported
detections of the Sgr stream, but also fitted the distance and the radial velocity of 
such detections.
The reddening maps of Schlegel, Finkbeiner, \& Davis (1998, 
hereafter SFD)  were used to select less extreme extinction windows along the predicted 
path of the stream (see Sec. 3.1).

\section{RESULTS}

\subsection{A SEARCH FOR THE SGR NORTHERN STREAM AT LOW GALACTIC LATITUDES}

The Sgr stream is oriented almost perpendicularly to the Galactic plane, which
it crosses  almost behind the Galactic
Center. For this reason, mapping the stream at low galactic latitudes ($b<20\arcdeg$) 
is very difficult because these regions are affected by severe differential reddening 
and high foreground contamination. Previous attempts to find the northern stream 
have failed (Mateo et al. 1998; Majewski et al. 1999). However, recent reports of a 
possible detection of this stream at 
$\sim$$60\arcdeg$ of the Sgr center (Yanny et al. 2000; Mart\'\i nez-Delgado et al. 
2001a: see Sec. 1) have provided a better determination of its orbit that motivates us 
for a new attempt to map it at lower galactic latitudes.  

Finding the Sgr stream at lower galactic latitudes could be difficult, unless some
predictions about the Sgr orbit be made using theoretical models.
Fig. 2 shows the position of the Sgr northern stream predicted by our model 
(dashed line; Mart\'\i nez-Delgado et al. 2001a) overplotted to the reddening map of 
SFD. This map shows how difficult  the detection of the stream at galactic latitudes 
$b<20\arcdeg$ is  using CMD techniques because of the severe differential reddening, which
would produce  the broadening of the CMD features of the stream's stellar population and
make them too diffuse to be detected above the Galactic foreground population (see 
Rosenberg et al. 2000 for some examples of reddening-distorted CMDs of globular 
clusters). Fortunately, we identify a low reddening window of $ 11\arcdeg\times 2\arcdeg$ 
situated at ($l,b$) $\sim$ (2$^\circ$, 13$^\circ$), that overlaps with the predicted 
position of the stream. We selected our target field in the position marked in 
Fig. 2 by the dot labeled  1 (SGR2+13\footnote{In the following, we will
name our target fields in the Sgr northern stream  SGR $l\pm b$, where ($l\pm b$) are 
the Galactic coordinates of the center of the field.}; see Table 2). The mean reddening 
of this field is $E(B-V)=0.240 \pm 0.017$. This 
value indicates that differential reddening should be sufficiently
limited for the presence of the Sgr stream to be detected from a CMD.

 Another problem is that this region is also affected by  high contamination 
from Galactic foreground stars. However, the CMD of the Sgr main body (Fig. 3a; 
see Sec. 2.1) shows how the MS turn-off of this galaxy lies in a gap in the distribution 
of contaminating Milky Way stars.  The predicted distance of the Sgr northern stream  
at SGR2+13 is $d_{o}\sim$ 25 kpc (Mart\'\i nez-Delgado et al. 2001a), which is very 
close to the  distance of the Sgr main body (24 $\pm$ 2 kpc; Mateo 1998). The difference 
in reddening between both fields is  E($B-V$)= 0.124 mag (see Table 2). This yields shifts 
in V  and $(B-R)$ at the Sgr MS turn-off of 0.43 and 0.2 magnitudes, respectively. 
The expected position for the Sgr MS turn-off in  SGR2+13 is hence very similar to 
that observed in Fig. 3a, indicating that the detection of the Sgr tidal stream in this 
field is feasible.

The CMD of SGR2+13 is shown in Fig. 3b, for comparison with the CMD of
the main body of Sgr (Fig. 3a). Although both diagrams show a similar foreground 
contamination pattern, no signature of the Sgr MS turn-off is observed in the CMD of 
our target field.  Since the path of Sgr in the l-direction is poorly constrained, 
the simplest explanation for this negative detection is 
that the Sgr northern stream is shifted by several degrees in galactic longitude ($l$) 
or it is narrower than expected, being placed our field SGR2+13 in the
outskirts of the stream. Alternatively, the 
surface brightness of this part of the northern stream might be too low to be detected.
During the revision process of the present paper, Majewski et al. (2003) have presented
new data of the Sgr stream, mapped with M giant stars, in the region of our field. A
detailed analysis of these data could help us to distinguish among the previous 
possibilities explaining our negative detection. Such a analysis will be performed
in a forthcoming paper, once the surface brightness of the Majewski et al. stream
be published.

\subsection{ DETECTION OF THE SGR NORTHERN STREAM AT b = 40$\arcdeg$ }

Ibata et al. (2001b) have suggested that a large fraction of the halo carbon stars 
discovered in their APM survey belong to the Sgr tidal stream, since they preferentially 
occur near the great circle of its orbit. Interestingly, this distribution includes 
the most conspicuous clump of carbon stars in the whole sky, centered at 
$\alpha \sim {\rm15^h}$, $\delta = -10\arcdeg$. This clump has a radius
of $\sim$$10\arcdeg$ and overlaps with the SDSS stream at the celestial equator. There are 
13  N-type carbon stars possibly associated with this clump, that is $\sim$1/3 of the 
carbon stars reported by Ibata et al. (2001b) as members of
Sgr. Whitelock et al. (1999) have estimated a total number of $\sim$100 N-type 
(i.e. intermediate-age population) carbon stars in the Sgr main body. Assuming the same 
stellar population (see Sec. 6) and scaling to the surface brightness in the main body 
of Sgr ($\Sigma_{V}\sim$= 25.4 mag arcsec$^{-2}$; Mateo 1998), we found that the expected mean 
surface brightness of the stream at the carbon star clump turns out to be  
$\Sigma_{V}\sim$ =27.2 $\pm$ 1.0 mag arcsec$^{-2}$. Therefore, a strong detection of the Sgr 
northern stream is expected in the position of this carbon star clump, assuming that it 
is associated with this galaxy.

To check this hypothesis, we have obtained a WFC field centered on the Ibata et al. 
carbon star clump (hereafter SGR352+42, labeled with number 3 in Fig. 2) together 
with a control field (see Table 2 for 
the coordinates of these fields). Because of the large projected angular extension of the 
Sgr stream in this region of the sky (at least $37.5\arcdeg \times 3.5\arcdeg$; Vivas et 
al. 2001), the control field was taken at $\sim 17\arcdeg$ from it. This would be far 
enough away to ensure that it is free from stream stars, but not too far for the 
foreground star distribution to change significantly.

 Fig. 4 shows the CMDs for the control field (panel {\it a}) and for SGR352+42 (panel 
{\it b}). A dense feature at $(B-R)\sim$ 0.8 and $22.5\leq$``$V$''$\leq 24$ is 
detected in the latter, which is not observed in the distribution of foreground Milky Way stars 
of panel {\it a}. Its color and shape correspond to what is expected for the upper MS of 
the Sgr dwarf galaxy, as  is shown in the CMD models plotted in Fig. 3a (see Sec. 2.2).  
A small number of sub-giant branch stars is also observed at ``$V$''$=21.5$ and 
$(B-R)\sim$ 1.0, indicating that a bright part of the Sgr stream has been detected, 
possibly associated with the apocenter of this galaxy.  Interestingly, our target field happens 
to be at the same position where Majewski et al. (1999) failed to detect the Sgr 
northern stream (their CMD was not deep enough to reach the MS turn-off).

To check the statistical significance of the proposed MS, we have
obtained star counts, in a box defined by $21.8\leq$``$V$''$\leq 23.0$, 
$0.6\leq (B-R)\leq 1.1$. The CMD was previously decontaminated from foreground stars 
using the procedure described in Gallart et al. (1996), assuming that the Galactic 
star population is the same in the target (Fig. 4b) and control fields (Fig. 4a). 
The net number of the proposed MS stars is found to be $N_{\rm MS}=240\pm 15$ 
(errors have been estimated assuming Poissonian statistics for the star counts), 
indicating with a high statistical significance level ($>99$\%) that the feature is real and
encouraging us to conclude that a real stellar system differentiated from the 
surrounding halo population has been observed. Scaling these
star numbers to those found in the CMD of the Sgr main body (see Sec. 2.1), 
the surface brightness magnitude of the new detection is $\Sigma$= 29.6 $\pm$ 0.2 mag arcsec$^{-2}$.

This Sgr tidal debris, situated  45$\arcdeg$ from the main body of the galaxy,
also  provides new data on the position, distance, and surface brightness of an up to 
date unknown part of the stream. The distance can be calculated from the magnitude of 
the MS turn-off, if its absolute magnitude and the interstellar reddening are known. 
We estimate that the magnitude of the MS turn-off (at the bluest point of the MS) is 
``$V_{o}$''$=22.45 \pm 0.20$. The interstellar extinction obtained from the reddening maps 
by SFD is $E(B-V)=0.075\pm 0.005$. The absolute
magnitude of the MS turn-off depends on the stellar population of the Sgr tidal stream. 
Under the assumption that
it is the same as in the main body of Sgr (see Sec. 6),  the CMD of the Sgr main body 
obtained by Layden \& Sarajedini (2000) can be used to set the distance modulus of the 
new tidal debris. This CMD suggests that the Sgr MS turn-off absolute magnitude is  
$M_{V}\sim 3.8\pm 0.1$. Using this value, the resulting distance modulus for the new tidal 
debris is $(m-M)_{o}=18.95\pm 0.22$, which corresponds to a distance $d_{o}=54 \pm 2$ kpc.

\subsection{ A DETECTION OF THE SGR NORTHERN STREAM AT b = 30$\arcdeg$ }

The coordinates of the new detection of the Sgr northern stream in SGR352+42 (see Sec. 3.2) 
were used to interpolate the position of a new target field during the campaign 
(SGR357+29; see Table 2 and Fig. 2, where is marked with label 2), 
situated $\sim$$10\arcdeg$ SE from SGR352+42. The CMD of this 
field is shown in Fig. 4c. Although SGR357+29 is affected by severe foreground contamination, 
the MS feature detected in SGR352+42 (Fig. 4b) is still visible  at a similar position in 
the CMD. We estimate the MS turn-off magnitude at  ``$V_{o}$''$\sim 22.08\pm 0.20$, 
that is +0.4 mag brighter than our former detection in SGR352+42. This turn-off magnitude 
variation is in agreement with the predictions of our theoretical model (see Sec. 4), in 
the sense that the heliocentric distance of the northern stream increases as we move away from the 
center of Sgr. Using the same procedure as in Sec. 3.2,  we calculate the distance
modulus of this new Sgr tidal fragment to be $(m-M)_{o}=18.28\pm 0.22$ ($d_{o}=45 \pm  2$ kpc).

Unfortunately, the lack of a useful control field at this galactic latitude
avoids to estimate the surface brightness of this part of the tidal stream  
detected in SGR357+29. However, scaling to the MS star density observed in 
their CMDs, the surface brightness of the stream in SGR357+29 (Fig 4c)  seems to 
be smaller than that observed in SGR352+42 (Fig 4b). This is consistent with our 
theoretical model,  predicting the surface brightness of the stream increasing close 
to the apocenter of Sgr. However, it is also possible that the new field is not 
completely centered on the stream, yielding a  less surface brightness detection.

\section {COMPARISON WITH THEORETICAL MODELS}

\subsection {THE NUMERICAL MODEL}

Our positive and negative detections of the Sgr northern stream reported in Sec. 3 
provide new observational data on its position and distance. This
information is extremely valuable for testing and refining our theoretical model, which 
is in turn fundamental for the reconstruction of the dynamical history of Sgr, and for 
constraining its total mass and dark matter content.

 With the purpose of updating our theoretical model, we have run $N$-body simulations of the 
Sgr plus Milky Way system.
The Milky Way model is a three-component system (spheroid + disk + halo) with
45~319 mass particles (see Fux 1997 for details). The halo represents the dark halo, 
while the spheroid accounts for
the nuclear bulge and  stellar halo. The spheroid density profile is inspired by the 
model of Sellwood \& Sanders (1988),
\begin{equation}
\rho_{\rm S}(s)=\frac{M_{\rm S}}{4\pi a^3 q_d I_{\infty}}\frac{(s/a)^p}{1+(s/a)^{p-q}},
\end{equation}
where
\begin{eqnarray}
&&s^2=R^2+z^2/q_d^2,\\
&&I_{\infty}=\frac{\pi}{p-q} \csc\left[\frac{(p+3)\pi}{q-p}\right],
\end{eqnarray}
$a=1.0$ kpc  (a knee radius), $q_d=0.5$ (the axis ratio of the density distribution), and
$M_{\rm S}=5\times 10^{10}$ $M_{\odot}$ (the spheroid integrated mass).
The values of $p$ and $q$ have been set to $-$1.8 and $-$3.3, respectively.
The mass density thus behaves as $s^{-1.8}$ for $s\ll a$, in
agreement with observations of the inner kiloparsecs of the Galaxy (Becklin \&
Neugebauer 1968; Matsumoto \etal 1982), and as $s^{-3.3}$ for $s\gg a$, similar to the radial number
density decrease of RR Lyrae stars (Preston, Schectman, \& Beers 1991) and 
of the globular clusters (Zinn 1985).

The stellar disk has a double exponential density
profile,
\begin{equation}
\rho_{\rm D}(R,z)=\frac{M_{\rm D}}{4\pi h^2_R h_z} \exp\left[-\frac{R}{h_R}-\frac{|z|}{h_z}\right],
\end{equation}
where $h_R=2.5$ kpc, $h_z=0.25$ kpc and $M_{\rm D}=4.6\times 10^{10}$
$M_{\odot}$ are the scale length, the scale height, and the
integrated mass, respectively. The Milky Way dark halo has an oblate
exponential density profile, with the same axis ratio, $q_d$, as the
spheroid component,
\begin{equation}
\rho_{\rm H}=\frac{M_{\rm H}}{8\pi b^3 q_d} \exp(-s/b),
\end{equation}
where $b=13.0$ kpc is the scale length and $M_{\rm H}=4.8\times 10^{11}$
 $M_{\odot}$ is the total dark mass. This density profile ensures an almost
flat rotation curve at large radii (between 15 and 80 kpc; see Fig. 1 of G\'omez-Flechoso et
al. 1999). All these components are
mildly truncated at large radii, multiplying their mass densities by
\begin{equation}
f(s)=\tanh\left[\frac{R_{\rm c}-s}{2\delta}\right],
\end{equation}
where $R_{\rm c}=70$ kpc is the truncation radius, $q_d=0.5$  is the
axis ratio of the density distribution and $\delta=5.0$ kpc provides 
the length over which $f(s)$ falls
from 1 to 0. Therefore, the densities vanish on the spheroid surface $s=R_{\rm c}$.
This galaxy model reproduces the Milky Way's characteristics (luminosity,
density profile, kinematics, etc; see Fux 1997 for details).

The Sgr satellite, represented by 8000 equal-mass particles, is a modified King
model (G\'omez-Flechoso \& Dom\'{\i}nguez-Tenreiro 2001) with 
initial mass $M_{\rm Sgr}=6.04\times 10^7$ $M_{\odot}$, core
radius $r_{\rm c}=0.3$ kpc, central velocity dispersion $\sigma_o=15.0$
km s$^{-1}$, and dimensionless central potential $W_{o}=3.0$. This model has been
built up in quasi-equilibrium with the potential of the Milky Way
model (see G\'omez-Flechoso \& Dom\'{\i}nguez-Tenreiro 2001 for details) and 
it has been evolved for roughly 5 Gyr. 
At this time, the satellite model fits  present-day Sgr characteristics.
The model initial conditions imply that the satellite is already
tidally truncated at the beginning of the simulation, and that  the
internal kinematics are in quasi-equilibrium with the satellite
density profile and the Milky Way tidal potential well. This means that
either the satellite was formed tidally truncated in the halo of
the Milky Way (Choi, Guhathakurta, \& Johnston 2002),
or it was accreted in the Milky Way and suffered a slow evolution which 
did not destroy the satellite in the early stages of its evolution. 
Our quasi-equilibrium model allows that Sgr has been 
orbiting the Milky Way more than 5 Gyrs, but we do not need to 
explicitly model the initial stages of the satellite evolution.
The scenario of a long living satellite is supported by the observations of the Sgr
stellar population and its tidal tail, which suggest that the
satellite has been orbiting the Milky Way for a few Gyr. Therefore, a
quasi-equilibrium King model, such as that described by G\'omez-Flechoso
\& Dom\'{\i}nguez-Tenreiro (2001) is justified. In contrast,
an arbitrary King model that does not take into account the
potential of the Milky Way will not survive for a long time and will not
reproduce the observations. 

Obviously, our results for Sgr mean
that the survival time of Sgr in the tidal potential of the Milky Way is 
longer than 5 Gyr. 
Our lack of knowledge concerning the potential of the Milky Way at earlier times
and its evolution history
makes it difficult to run realistic simulations of the Sgr + Milky Way system
for longer times and it is out of the scope of this paper.
To properly simulate the Sgr + Milky Way system the Sgr satellite and the 
Milky Way have to be modeled. The evolution of the Milky Way density and
potential is not well known because depends on the cosmological model, since it
describes the accretion rate of satellites, that increases the mass of the main
galaxy, and the redshifts at which the accretions are more important. As 
modeling the evolution the Milky Way is difficult and is not one of the aims
of this paper, we have restricted the simulations to the last 5-6 Gyrs. 
During this period of time all the realistic cosmological models agree on the fact 
that the potential of the Milky Way has remained substantially constant. In 
this way, we avoid the uncertainty introduced in the evolution of Sgr by a poor 
knowledge of the Milky Way evolution. Any other simulations modeling the evolution
of the Sgr + Milky Way system for longer periods of time (more than 5-6 Gyrs) 
should consider the cosmological
evolution of the Milky Way potential that will also modify the evolution of the Sgr
satellite. Using cosmological models, Helmi
et al. (2003) have obtained the accretion rates in the main halo as a function 
of the halo radius at different redshifts. These models shows that the mass 
accretion in the inner 100 kpc in negligible for the last 6 Gyrs, however this
mass accretion could be important at early times.

The Sgr orbit has been selected to match the observations of Sgr main
body, that is, the center of Sgr and the tidal tail close to it.
The other observations of the Sgr stream are not taken as inputs to
selected the best Sgr model, but they are confirmed as part of Sgr stream
using the model.
The pericenter and apocenter of the best-fitting
orbit are 12 and 60 kpc, respectively, and the anomalistic period
is $\sim$$0.74$ Gyr. The present position and velocity of the
satellite model (after 5 Gyrs orbiting the Milky Way) are
$(X,Y,Z)=(16.0,2.0,-5.9)$ kpc and $(U,V,W)=(241,16,240)$ km
s$^{-1}$, reproducing the Sgr observations. Other characteristics
(internal velocity dispersion, size, density, etc.) are also
fulfilled (see Table 3 for details). Henceforth, 
we will refer to this snapshot of the
simulation.

\subsection {FITTING TO THE OBSERVATIONS OF THE SGR TIDAL STREAM }

Fig. 5 and Fig. 6a show the projection on the sky and the heliocentric 
distance of the tidal stream model 
respectively ({\it black dots}).  
In both figures, the corresponding observational data of the main body of Sgr ({\it 
red square}; Ibata \etal 1994) and its southern tidal stream ({\it solid 
purple line}; Mateo \etal 1998) are
also plotted, as well as the detection of this stream by 
Majewski \etal (1999; {\it open blue square}). Other observations related to 
the Sgr southern stream are the globular cluster Pal 12 ({\it green cross}; 
Dinescu et al. 2000; Mart\'\i nez-Delgado et al. 2002b) and the A-type star 
stream reported by the
SDSS stream in the South Galactic Pole (Yanny et al. 2000; Newberg et 
al. 2002) and the detection of Sgr tidal debris in the field SA57 by Dinescu et al. (2002) 
({\it solid blue diamond}). Fig. 5 also shows a detail of the model in the region of 
the Sgr northern stream. This includes the reported detections of A-type 
stars (Yanny \etal 2000) and the RR Lyr stars (Ivezic \etal 2000) from 
the SDSS stream ({\it shaded areas}), the first detection of the stream using a CMD by Mart\'\i nez-Delgado et al. 2001a 
({\it purple triangle}) and the carbon stars clump reported by Ibata \etal (2001b: 
see Sec 3.2). The position of our new positive ({\it filled purple circles}) and 
negative ({\it open purple circles}) detections of the Sgr northern stream (including the control field)
reported in Sec. 3 are also shown in this panel. 

During the revision process of this paper, Majeswki et al. (2003) have presented the first all-sky view of the Sgr tidal stream using M giant star tracers selected
from the 2MASS. Our model agrees quite well with the projected orbit of the Sgr stream reported by this study. The only disagreement between the model and these observations corresponds to the 
region situated at $60 <$ RA $< 120$ and $60 <$ Dec $< 120$. As it is discussed in Sec. 6 (see also Fig. 15), this 
part of the stream is the oldest one, since it has been unbound more that 5 Gyrs ago. This region also 
corresponds to one of the less populated
and furthest part of the Sgr stream. The reasons of the disagreement could be 
either a possible
evolution of the Milky Way potential in this region during the last 5 Gyrs or the 
uncertainty on the observational proper motions of Sgr, since small variations of the proper
motions produce large differences in the position of the stream at large distances
from the Sgr center. A more detail analysis of these hypothesis will be addressed in a future paper.

Our comparison of the distance estimates for these detections with the prediction of
our theoretical model (Fig. 6a) shows as the most accurate distances for the stream come from those
studies based on the main-sequence fitting of the CMD or on RR Lyraes for which complete light curves are available (Vivas et al. 2001). Both methods also provide a reliable estimated in a well established distance scale.  For the stellar 
overdensities reported by SDSS, errors in the distance have been calculated assuming $\pm 0.5$ and $\pm0.25$ mag errors for A-type and RR Lyrae stars, respectively. The distance to the carbon stars
reported by Ibata et al. (2001b) close to the Sgr apocenter suggest that the stream could 
spread out  to at least 80 kpc in the same direction of the sky. However, the errors in the distances of carbon stars by Ibata et al. are large (about 25$\%$) and some carbon stars could be enveloped in dusty shells causing their distances to be overestimated. 
 
The comparison of the available kinematic data of the Sgr tidal stream with the heliocentric radial
velocities predicted by of our model is given in Fig 6b. The lack of a systematic kinematic survey and the heterogenous nature of the radial velocity estimates (specially because they come from different stellar tracers of the stream) are a strong limitation in the comparison of the observational data with our model. In spite of this, the agreement of these observations with our model is rather good. In addition of the  detections labeled in Fig 5, we also plotted the radial velocities reported by Yanny, Newberg et al. (2003) ( {\it blue shaded area} ) in the Sgr southern stream.  
 The radial velocities of the carbon star clump (Totten \& Irwin 1998; Totten,  Irwin, \& Whitelock 2000) are also very similar to those of the stream's red giants reported 
by Dohm-Palmer et al. (2001) in this region of the sky. Both samples show an striking agreement in radial velocities for distances between 40 and 80 kpc, with a mean value of $V_{r}=56 \pm 32$ km s$^{-1}$, that is also in very good agreement with the prediction of our model. This suggests that both samples could belong to the same coherent structure belonging to the Sgr northern stream.

In summary, we find a good agreement of our new model with the available observations of the 
Sgr tidal stream. However, this model is not the only one that reproduces 
the present observations. A more massive satellite model in equilibrium with the environment 
will be able to survive orbiting the Milky Way halo for a longer time before reproducing 
the degree of disruption currently shown by Sgr, but in this case the mass in the tidal stream 
(that is, the material disrupted from the satellite) would be larger. For this reason, 
observations of the luminosity and kinematics of the stream are important in choosing 
the most accurate Sgr model.

\subsection{THE CLOSER WRAP OF THE SGR STREAM  }

 Our model of Sgr predicts that, after the apocenter region, the Sgr northern stream starts to fall on the Galactic disk, in good agreement with previous theoretical models (Ibata et al. 2001b). This loop of
the Sgr leading tail is better seen in the XZ-plane projection of the Sgr stream model  with respect to the galactic center (Fig. 7). The Sun's coordinates are $(X,Y,Z)_{\odot}=(-8.5,0.0,0.0)$ kpc and Sgr center is placed at $(X,Y,Z)_{Sgr}=(16,2,-5.9)$ kpc.  This prediction has recently received strong observational support from the Sgr orbit traced by its M giant population  (Majewski et al. 2003).

Fig. 8 shows a detail of the projection on the sky and the heliocentric distance predicted 
by our model for this part of the Sgr northern stream.
The distance range of the stream stars is indicated in different colors. 
The model suggests the presence of nearby Sgr tidal material sparse in a huge area of the sky. 
This could be related with the detection of a low-surface brightness, blurred distribution of M 
giants close to the North Galactic Pole by  Majewski et al. (2003). The QUEST survey (Vivas  \& Zinn 2002) reported a clump of 21 
RR Lyraes located at 19 kpc from the Sun in this region of the sky ({\it filled purple square}), but they do not report it as Sgr detection and associate it with a hitherto unknown tidally disrupted dwarf galaxy.  Newberg et al. (2002) report the presence 
of MS stars associated to this RR Lyrae population in a SDSS slice (named S297+63-20.0) situated at the same position ({\it open rectangle}).
However, they also claim these stars to be associated to an unknown tidal stream or other diffuse concentration of stars in the halo. Fig. 8 shows that both detections are in agreement in position and distance with 
the prediction of our theoretical model for the trailing arm of the Sgr tidal stream. 

 Our model also suggests that the leading and trailing streams overlap at  $\sim$ 20 kpc from the Sun, 
in the direction of the line-of-sight plotted with a solid line in Fig. 7. Because of the presence 
of stream material from these two tails, an enhancement of the surface brightness density for this 
nearby wrap of the stream is expected towards this direction. A possible detection of this nearby 
tidal debris shell is reported by Martinez-Delgado (2003) from a deep CMD in this area. A kinematic 
study of this area might reveal different peaks in the radial velocities distribution of the Sgr 
stars in  this region (see Zinn 2003), what would be very valuable to constrain the models in this 
part of the northern stream.

As it is also noted by Majewski et al. (2003),  the leading arm passes through the Galactic plane 
in the proximity of the solar neighborhood. Our distance measurements of the MS stars observed in the Sgr 
northern stream (see Sec. 3) are accurate enough to constrain the model of the Sgr orbit
and it provides a good estimate of the position of the stream when it crosses the Galactic plane. Fig. 9a
displays a detail of the XZ-plane projection in Galactic coordinates of our model. It shows the 
leading arm falling almost perpendicular to the Galactic plane in the solar neighborhood. The circle 
centered in the Sun marks the outer limit in heliocentric distance ($\sim$ 4 kpc) of the local halo 
sample from the Chibas \& Beers (2000) catalogue. Our model predicts than some
stream stars could contaminate the local sample and could be identified as
halo proper motion group members in the solar neighborhood\footnote{ A comparison of the model 
predictions with known proper motion groups observed in the solar neighborhood is out of the scope of 
this work and will be discussed in a forthcoming paper.}.
Fig. 9b plotted the spatial distribution of Sgr stream in the XY-plane given by the model, restricted to 
the points of the simulation at Galactic azimuthal distances $|Z|< 4$ kpc. This shows that the Sgr stars should be detected in the direction l$\sim$ 320$\arcdeg$. Interestingly, Kundu et al. (2002) reported a group of giants possible associated to a diffuse part of the leading tidal arm of Sgr in this direction. These stars are marked with open purple 
triangles in Fig. 9. The comparison of the radial velocities of these stars ({\it open triangles}) obtained by Kundu et al. (2002) with our model ({\it filled circles}) for this region of the Sgr stream is given in Fig. 10. The agreement of these data with the model prediction is very good, yielding to
confirm the Kundu et al's sample is an observational evidence
on the presence of the debris material of the Sgr leading arm in the proximity of the solar neighbourhood.

\subsection{ MULTIPLE WRAPS OF THE SGR STREAM?}

 In addition to the main Sgr northern stream studied in this paper (formed by material that
was lost in recent pericentric passages; see Sec. 6), some theoretical models  (Helmi \& White 
2001; Ibata et al. 2001b)  predict multiple wraps of debris material located at larger 
distances (between 80 and 100 kpc) in the region of the SDSS stream. This stream would 
have been  formed by older material that became unbound more than 7 Gyr ago.

However, our theoretical model (see Sec. 4) cannot predict the presence of this early 
wraps of Sgr at this distance, as is shown in Fig. 6, because we have restricted our simulations
to the last 6 Gyrs, and these wraps are expected to be formed earlier. 
The energy of the present orbit of Sgr is not high enough to have an
apocenter at 80 kpc. The Helmi \& White (2001) and Ibata et al. (2001b) models 
predict this stream after assuming that Sgr was in a more external orbit that 
dissipated energy up to 
reach the present one. They use a satellite model that is not
in equilibrium with the Milky Way potential. Therefore, when the satellite model
is introduced in the Galaxy potential suffers a strong (and spurious) kick that
forms the dense tail predicted at large distances.
Additionally, their simulations try to reproduce the Sgr disruption process in the last 12-14 Gyrs. 
In these long term simulations, the possible evolution of the external halo of the Milky Way 
(Helmi et al. 2003), that could affect the orbit and the disruption process of the satellite, 
has to be taken into account. It introduces new sources of uncertainty in the numerical
results which are difficult to estimate and which has not been taken into account by the
forementioned authors.
In our model, we assume more realistic initial conditions for the satellite
model, since we use quasi-equilibrium King models (G\'omez-Flechoso \& 
Dom\'{\i}nguez-Tenreiro 2001) which take into account the Milky Way potential. 
In this scenario, the satellite does not suffer the spurious initial kick.
As a consequence, the satellite mass loss of our model is slower.
In fact, the stream of our model only traces the last orbit, which corresponds to 
the present orbit of Sgr with apocenter at $\sim 60$ kpc.
Moreover, as we restricted our simulations to the last 6 Gyrs, we skip the problem of a
possible evolution of the Milky Way potential, since it is supposed to remain 
almost constant during the last Gyrs.
Therefore, our simulated streams do not show any trace of previous orbits contrary to 
the forementioned models. In addition, a model with traces of earlier orbits would represent a 
more disrupted satellite and would therefore  show  tidal tails in the Sgr 
neighborhood brighter than those observed.

Observations of this predicted multiple wraps of the Sgr tidal stream would thus provide a strong constraint to the theoretical models. Dohm-Palmer et al. (2001) reported evidence for additional structures at different distances in the region of the SDSS stream from a survey of giants
in the SDSS stream. Although the errors in the distances of this sample (about 25$\%$) prevent an accurate 
estimate of the depth along the line of sight of the Sgr tidal stream, they
estimated that this stream could be spread out to at least 80 kpc in this direction of the sky. However, the search for overdensities of faint blue stars in the SDSS using the technique described in Yanny et al. (2000) has failed in detecting the multiple wrap structure of the Sgr northern stream  predicted by these models.
Since the surface brightness of the unbound material of Sgr decreases with the time,  a possibility is that this technique fail in finding this fainter wrap of tidal debris originating in earlier passages of Sgr. The RR Lyrae  surveys in the SDSS stream  have shown evidence of the presence of a long and relatively
thin clump of RR Lyraes situated at 50 kpc (Ivezic et al. 2000; Vivas et al.
2001), in good agreement with the model predictions of the Sgr northern stream.
However, the mentioned models predict a second peak in the distribution of RR Lyrae stars at $V$ = 20.5 mag, corresponding to the distances of  this further wrap (Helmi \& White 2001). The deeper survey by Ivezic et al. (2000) have not found any evidence of this further clump of RR Lyrae which would be related to the second wrap of the Sgr tidal stream. Helmi \& White (2001) estimate that the expected number of  RR Lyrae for the second tidal stream in this survey would be $\sim 2\pm 1$ 
and  conclude that the failure to detect it from these data is not conclusive.  A deeper RR Lyrae 
survey in a significantly larger area is necessary to definitively accept or 
reject the presence of a distant wrap in this direction of the sky. There is no evidence of this distance lump from the 2MASS survey (Majewski et al. 2003). However, 
it is important to note that the stream material from this distant filament 
corresponds to material unbound more than 5 Gyr ago (see Sec. 6), and could be predominantly 
composed by metal-poor, old stars. Therefore, it could remain undetected to the 2MASS survey, that 
is almost blind to metal-poor, old tidal streams in the halo. 

In summary, the detection of this further Sgr stream structure from current photometric surveys is very hard. Kinematics and proper motion studies
  and the future generation of astrometric satellites 
(GAIA, SIM, FAME) could give an important boost in this issue.

\subsection {THE ARNOLD \& GILMORE'S HALO PROPER MOTION GROUP: ANOTHER SGR TIDAL DEBRIS }

In the last decade, different studies reported the presence of isolated proper motion groups in the Galactic halo that are considered as evidence of mergers during the evolution of the Milky Way. Arnold \& Gilmore (1992) found a group of four A-stars at 30 kpc of the Sun with a velocity dispersion of less 
than 12 km s$^{-1}$, concluding they belong to a halo moving group, associated to "the stellar remnants of
a recently disrupted halo cluster".   

The comparison of the position, distance and radial velocity with our Sgr model ({\it blue solid triangle} in Fig. 5 and Fig. 6) reveals that this group
of stars could be actually associated to the Sgr tidal stream. The position of F826 field 
($\alpha$=01$^{h}$ 00$^{m}$, $\delta$=0$\arcdeg$) is close to the border of the region where the SDSS reported a prominent A-stars clump associated to the Sgr southern tidal stream (named S167-54-21.5 in Newberg et al. 2002). The  mean distance ($\sim$ 30 kpc) and radial velocity ($V_{r}$= -143 $\pm$ 6 km s$^{-1}$) for Arnold \& Gilmore's group is also in very good agreement with the predictions of our model. From this result, we conclude that this halo proper motion group is actually
a tidal debris of Sgr. Interestingly, this proper motion group can be then considered like the first detection of the Sgr tidal stream, two years before the discovery of the main body of this galaxy by Ibata et al. (1994).

\section {CONSTRAINING THE DARK MATTER HALO OF THE MILKY WAY}

In the previous section we have explained the Milky Way model used to 
reproduce Sgr and its stream. We have mentioned how the luminosity
and the matter content of the stream could give us information about the
initial mass of Sgr. Additionally, the shape and the width of the stream
provide information about the Milky Way potential. 
The potential is the physical quantity that determines the satellite orbit.
In order to infer the shape of the Milky Way potential, 
we will need to know the distance and the velocity of the satellite along
its orbit. This cannot be possible to obtain from the observational data 
of a satellite, unless it has a tidal stream. As the tidal stream follows 
roughly the same orbit as the satellite, if we measure the distances and the
velocities of the stars in the stream, we can infer the kinematics and the
shape of the satellite orbit. Using these data, we can obtain the 
halo potential that better reproduces all the characteristics of
the satellite orbit.

Our best Sgr model (described in Sec. 4) has been obtained using 
an oblate density distribution ($q_d=0.5$) for the dark matter halo. However, 
the potential produced by this density distribution is more spherical  
and its flatness varies with the distance to the center of the Milky Way, as it
is shown in Fig. 11. As it can be seen, the potential flatness is 
important only for distances to the Milky Way center smaller than 10 kpc 
and at these distances the flatness is mainly produces by the disk potential. 
For the average distance of the Sgr orbit the potential flatness is roughly
$q_p \sim 0.85$.

This result is consistent with the cosmological Cold Dark Matter models of structure formation. In 
these models (e.g. Dubinski 1994) the density oblateness of the dark halos is $\sim 0.5$.
Our result also agrees with previous studies that infer an almost spherical 
potential for the Milky Way (Olling \& Merrifield 2000; see also Chen et al. 
2001). We wish to comment that the spatial 
distribution of the Sgr stream and the velocity of its stars can 
give us information only about the potential of the Milky Way, but not about its 
mass (or density) distribution.  In fact, different density distributions can 
reproduce similar potential
wells. This could explain the pretended disagreement between our results and those
by Ibata et al. (2001b), who postulate
an almost spherical density distribution for the Milky Way halo, which also corresponds to an
almost spherical potential well.

The torque exerted by the Milky Way potential in the Sgr orbit is another 
reason given by these authors for inferring a spherical halo. 
An oblate potential produces
a precession in the orbit that should be seen in the Sgr stream if two 
fragments of the stream belonging to two different orbits coincide in
the same region of the sky. Ibata et al. (2001b) argue that, since such a 
precession is not observed in the Sgr stream, which is seen as a single great circle, 
 the Milky Way halo must be almost spherical. However, in our simulations 
the tidal debris obtained is consistent with a single great circle distribution
on the sky (see Fig. 5). The first reason is that the flatness of 
the Milky Way potential is small in almost all positions along the Sgr orbit. 
The second reason is that the full stream mapped covers only one orbit
of Sgr. Therefore, the precession produced on different orbits cannot be 
observed, because no signal of previous orbits are appreciated on the sky.

We have run various simulations with different flatness parameters for the 
Milky Way halo. Our conclusion is that the variation of the halo flatness (with
the same total halo mass) produces a Sgr orbit with different peri and apocenter.
The radial velocity of the satellite at a given position in the orbit also 
depends on the halo flatness. However, in the case of Sgr, the projection of the
orbit on the sky is almost independent of the halo shape. Fig. 12 
plotted this projection for various halo density flatness. The Sgr orbit in all
the cases has been selected to fit the present position and velocity of Sgr.
As can be seen, almost all the models fit the confirmed detection of Sgr. 
Therefore, the projected orbit is not enough to distinguish among the feasible
values of the halo oblateness. In order to infer the shape of the halo potential
other observational data of the Sgr stream, as the heliocentric distance and the
radial velocity, are needed. In Fig. 13a and 13b we have plotted the few
observational distances and radial velocities, respectively, obtained for the 
Sgr stream. The results for the halo models with different flatness are also
plotted in the same figures. In Fig. 13a it can be seen that a more spherical
halo produces a Sgr orbit having the apocenter further away, and vice versa. 
The radial velocity at a given point in the orbit also depends on the halo 
potential (Fig. 13b), for spherical halos the range of radial velocities is 
larger than for flat potentials. The range of halo oblateness that is consistent
with the observational data of the Sgr stream corresponds to a density flatness,
$q_d$, between 0.4 and 0.6. It translates into potential flatness, $q_p$, between
0.8 and 0.9 for the average distance of the Sgr satellite in its orbit, as can 
be seen in Fig. 11. However, more observational data of the radial velocity and
the heliocentric distance of the Sgr stream are needed to confirm these results.

The spread of the Sgr stream is also a source of information about the 
characteristics of the halo. A lumpy halo produces a larger spread 
in the stream because the collisions between the stream stars and
the halo clumps involve larger energy transfers.  In contrast, a smooth halo 
keeps the stream narrow. In our simulations we have used an $N$-body halo.
 This is not a completely 
smooth halo, but  neither is it  very clumpy. Better observational data
on the width of the Sgr stream and higher resolution simulations are 
needed in order to further constrain the clumpiness of the Milky Way halo.

\section{THE STELLAR POPULATION OF THE SGR NORTHERN STREAM}

The study of the stellar population variation along the Sgr tidal stream can 
provide constraints on theoretical models of its tidal disruption, the epoch of its accretion in the Galactic halo, and helps us to understand the role of this process in its internal evolution.

It is well known that the Sgr central region has experienced a 
complex star formation history. Layden \& Sarajedini (2000) found that the  stellar 
population there has been built up through
 main episodes of ages 11, 5 and 0.5--3 Gyr and [Fe/H] values of $-$1.3, 
 $-$0.7, and $-$0.4, respectively. Cacciari et al. (2002) found
the existence of two clear components of red giant branch stars, with  [Fe/H] peaking at 
about $-$0.6 and $-$1.55 and a possible hint of a component at about $-$2.1. 
They also 
reported the presence of corresponding RR Lyrae populations with metallicities [Fe/H] $-$1.2 
and $-$2.1 (the metal-rich Sgr population at [Fe/H] = $-$0.6 does not produce RR 
Lyraes, as expected by the models, see Layden 1994) and small number of blue horizontal branch stars
(Monaco et al. 2003). From the 2MASS infrared photometry of 
a Sgr central field, Cole (2001) determined a mean metallicity for the main stellar 
population of [Fe/H] = $-0.5\pm0.2$. He also found the traces of the minor metal-poor 
population reported in  former studies.

There is evidence  of the presence of stellar populations of different ages in the Sgr 
tidal stream too. The detection of a significant RR Lyrae population associated with the 
northern stream (Ivezic et al. 2000; Vivas et al. 2001) is a strong indication of the presence 
of an old, metal-poor population. The identification of  bright carbon stars (Ibata et al. 2001b) and a bright M giant population (Majewski et al. 2003) along the Sgr  stream   points to the presence of a metal-rich
intermediate-age population along the stream too. This is consistent with the CMD morphology of the 
SDSS stream by Newberg et al. (2002), which shows a conspicuous red clump of helium-burning stars.

Fig. 14 shows the CMD for the Sgr northern stream in the SGR352+42 field, situated in a region close to the apocenter of Sgr. Isochrones from the Padova library (Bertelli et al. 1994) of $Z=0.001$ ([Fe/H]
=  $-1.3$) and age 11 Gyr ({\it solid line}) and  $Z=0.004$ ([Fe/H] = $-$0.7) 
and age 5 Gyr ({\it dashed line}) are overplotted. This figure shows that the 
stellar population of this part  of the Sgr northern stream is consistent with  star formation episodes with  ages 11 and 5 Gyr and [Fe/H] values of $-$1.3
and  $-$0.7, respectively, as found by Layden \& Sarajedini (2000) in the Sgr main body. Although it is not possible to conclude from the CMD 
whether both stellar populations are positively represented in this region of the stream, the CMD morphology suggests that this apocenter region could share the complex star formation history of the main body of the galaxy.

The main question is whether this complex stellar population could be the typical population of Sgr when it was accreted by the Milky Way and started to be disrupted by the Milky Way tidal potential. This would mean that Sgr was accreted by our Galaxy when the halo was already formed. Sgr would therefore have had  time to evolve in isolation  and  produce 
an enriched population before its accretion and disruption by Milky Way tidal forces. In this case, if the halo is formed by the accretion of satellite galaxies, the progenitor satellite galaxies would
have been  different from Sgr. The alternative hypothesis is that the Sgr star formation history could be the result of different star formation burst, possibly related with its different pericenter passages. In this case, a variation in the stellar population correlated with its  distance to the Sgr center is expected. The reason is the distance between a region of the Sgr stream and the center of the galaxy is related with  the time when this stream material was disrupted, the stream  stars situated at larger distances from the satellite center becoming unbound at earlier epochs. Therefore, the stars at the greatest distances are 
representative of the population of the outskirts of Sgr 
when the tidal stream started to be formed. In the disruption process, 
the stars which became unbound are those with higher energy (i.e. placed in the external part of Sgr).
Therefore, the stars in a given region of the tidal stream are representative of the external
part of the satellite population at the time of their disruption. 
If a new population of stars is formed at the center of the dwarf (where the satellite density is higher), it
will not be reflect in the tidal tail immediately, since these new stars have 
to migrate to the Sgr outer regions before being disrupted.

The study of the stellar population gradient along the stream is therefore very important to constrain the 
epoch of accretion of this galaxy and to understand if Sgr could be an actually "building-block" of 
the Galactic halo.  Some insights on the presence of a stellar population gradient can be obtained from the metallicity estimate of Sgr stars in different part of the stream. Dohm-Palmer et al. (2001) reported
spectroscopic abundances for red giant stars from the Sgr northern stream and found
a mean value of [Fe/H] = -1.5. Mart\'\i nez-Delgado et al. (2002b) find a mean 
metallicity of [Fe/H]=  -1.2 from the CMD of the remnants of the Sgr detected 
around the globular cluster Pal 12. Both estimates are consistent with the metallicity of the main
metal-poor component observed in the main body of the galaxy. However, some studies report the
presence of a more metal-poor old stars ([Fe/H] $\sim -2.5$), such as those that compose the stellar population of the outer halo.
Mart\'\i nez-Delgado (2003) report a mean metallicity for the stream stars of [Fe/H] $\lesssim -2.0$
in a field close to the globular cluster NGC 4147. This is consistent with the metallicity range ($-1.4 < [Fe/H]<-2.0$) obtained from spectroscopy measurement of RR Lyrae by Vivas \& Zinn (2003).

Fig. 15 shows these mean metallicities overplotted in the  aitoff sky projection of our  Sgr stream model. In this figure, the black particles are still bound to the Sgr galaxy; the
yellow particles got unbound during the last Gyr; the
green particles became unbound between 1.0 and 2.0 Gyr ago; 
the blue particles, between 2.0 and 3.0 Gyr ago; the
purple particles, between 3.0 and 5.0 Gyr ago; and the red particles,
more than 5.0 Gyr ago. In spite of the small number of  metallicity measurements,  the different metallicity values of the Sgr stars in different regions of the stream  shows the existence of a metallicity gradient along the Sgr tidal stream. These data also suggest that the wraps of  the tidal stream situated at larger distances of the Sgr main body (that is, formed in earlier mass loss events of the galaxy) could be composed by predominantly old, metal-poor stars, with a metallicity similar to those observed in the Galactic halo population. A systematic study of the stellar population gradient along the Sgr stream is necessary to confirm or reject this result.

\section {CONCLUSIONS}

We report the detection of two new  wraps of tidal debris of the northern stream of 
the Sgr dwarf galaxy located at  45$\arcdeg$ and 55$\arcdeg$ from the center of the
galaxy. 
The first one overlaps with the prominent clump of carbon stars discovered by Ibata et al. (2001b).
The derived distances for these debris are 45.3 kpc and 53.7 kpc respectively and both
detections are also 
strong observational evidence that the tidal stream discovered by the SDSS is tidally stripped material from the Sgr dwarf and support the idea that the
Sgr tidal stream completely enwraps the Milky Way in an almost polar orbit. 

To confirm these observations, we have run numerical simulations of the Sgr dwarf 
satellite plus the Milky Way. 
The Milky Way model is a numerical three-component system, whose parameters has been selected 
in agreement with the observations of our Galaxy.
The Sgr model has been built up in equilibrium
with the tidal forces of the Milky Way. This avoids the spurious disruption that routinely
show up when physically unrealistic models are used.
The initial conditions of the orbit were chosen to fit at the
end of the simulation the present observations of the Sgr velocity and position. The model 
presents at the end of the simulation a long tidal tail that traces the last orbit. The positions 
and the distances obtained in the new detections mentioned above are fully consistent with 
this model of Sgr.

 This model is also in good agreement with the available observations of the Sgr tidal stream. The comparison of our model with the positions and distances of two non-identified halo overdensities discovered by the Sloan Digitized Sky Survey and the QUEST survey shows that they are actually associated to the trailing arm of this stream. In addition, we identify the proper motion group discovered by Arnold \& Gilmore (1992) as a piece of this northern stream, and could be considered like the first detection of the Sgr tidal stream.

We also present a method for estimating the shape of the Milky Way halo potential using numerical simulations.
From our simulations we obtain an oblateness of the Milky Way dark halo potential of 0.85, using the current database of distance and radial velocities
of the Sgr stream. This result is consistent with the cosmological models of structure formation.

The CDM of the apocenter of Sgr shows that this region of the stream shares the complex star formation history observed in the main body of the galaxy. We present the first evidence for a gradient in the stellar population along the stream, possibly correlated with its different pericenter passages.

We are grateful to  E. K. Grebel and S. Majewski for several fruitful suggestions. 
DMD is grateful to M. Mateo for many useful comments and for his training in the real-time 
photometry techniques used in this paper. We also are grateful to R. Fux for providing the
program to build the Milky Way numerical model. This paper is based on observations made with 
the 2.5 m INT, operated by the Isaac Newton Group in the Spanish Observatorio del Roque 
de Los Muchachos of the Instituto de Astrof\'\i sica de Canarias.

\newpage

\begin{deluxetable}{cccc}
\tablenum{1}
\tablewidth{400pt}
\tablecaption{Journal of observations
\label{journal}}
\tablehead{
\colhead{Date} & \colhead{Field}  & 
 \colhead{Filter} & \colhead{Total exposure time (s)}}
\startdata

2001 Jun 22 & CONTROL  & $B$ &  8400 \nl
2001 Jun 22 & CONTROL  & $R$ &  7200 \nl
2001 Jun 25 & SGR2+13  & $B$ &  4900 \nl
2001 Jun 25 & SGR2+13  & $R$ &  3600 \nl
2001 Jun 26  & SGR352+42  & $B$ &  3600 \nl
2001 Jun 26  & SGR352+42  & $R$ &  2400 \nl
2001 Jun 28  & SGR357+29  & $B$ & 5400  \nl
2001 Jun 28  & SGR357+29  & $R$ & 5400  \nl

\enddata
\end{deluxetable}

\newpage

\begin{deluxetable}{lccccc}
\tablenum{2}
\tablewidth{400pt}
\tablecaption{Target fields
\label{Target_fields}}
\tablehead{
\colhead{Field} & \colhead{R.A. (J2000.0)} & \colhead{Dec} & \colhead{\it l} &
\colhead{\it b} & \colhead{$E(B-V)$}}
\startdata
SGR2+13\tablenotemark{ a}    & $17^{\rm h} 02^{\rm m} 43\fs3$ & $-19\arcdeg 45\arcmin 00\arcsec$ &   2.25 &  13.21 & $0.240\pm0.017$ \nl
SGR352+42\tablenotemark{ b}  & $15^{\rm h} 11^{\rm m} 06\fs8$ & $-07\arcdeg 31\arcmin 22.4\arcsec$ & 352.29 &  41.59 & $0.075\pm0.005$ \nl
 SGR357+29\tablenotemark{ c}  & $15^{\rm h} 57^{\rm m} 20\fs6$ & $-13\arcdeg 50\arcmin 20.0\arcsec$ & 356.60 &  29.08 & $0.157\pm0.009$ \nl
Sgr & $18^{\rm h} 33^{\rm m} 26\fs0$ & $-40\arcdeg 57\arcmin 57.0\arcsec$ & 353.89 & $-$14.25 & $0.116\pm0.005$ \nl
Control & $16^{\rm h} 11^{\rm m} 04\fs0$ &  $+14\arcdeg 57\arcmin 29\arcsec$ & 28.7 & 42.2 & $0.038\pm0.005$ \nl
\enddata
\tablenotetext{a}{Label 1 in Fig. 2}
\tablenotetext{b}{Label 3 in Fig. 2}
\tablenotetext{c}{Label 2 in Fig. 2}
\end{deluxetable}

\newpage

\setcounter{table}{2}

\begin{table}[h!]
\caption{Model de Sgr
\tablenum{3}
\label{tablemodel}}
\begin{tabular}{lll}
\\
\hline\hline
\multicolumn{3}{l}{\sc Sgr initial model}\\
King model parameters & $W_o$ & 3.0\\
           & $r_o$ & 0.3 kpc\\
       & $\sigma_o$ & 15 km/s\\
       & $M_o$ & $6.04\times 10^7$ M$_{\odot}$\\
       &&\\
\multicolumn{3}{l}{{\sc Sgr final model} (after $\sim 5$ Gyr)}\\
Bound mass & $M_{\rm Sgr}$ & $7\times 10^6$ M$_{\odot}$\\
Tidal radius & $r_{\rm t}$  & $1.4$ kpc\\
Half mass radius & $r_{\rm h}$ & $0.53$ kpc\\
Central luminosity\tablenotemark{a} & $\mu_o$ & 25.3 mag \\
Final position & $(X,Y,Z)$ & $(16, 2, -5.9)$ kpc\\
Final velocity & $(V_x,V_y,V_z)$ & $(240,16,248)$ km s$^{-1}$\\
\hline\hline
\tablenotetext{a}{We have considered $M/L=5$}

\end{tabular}
\end{table}

\newpage

\begin{figure}
\plotone{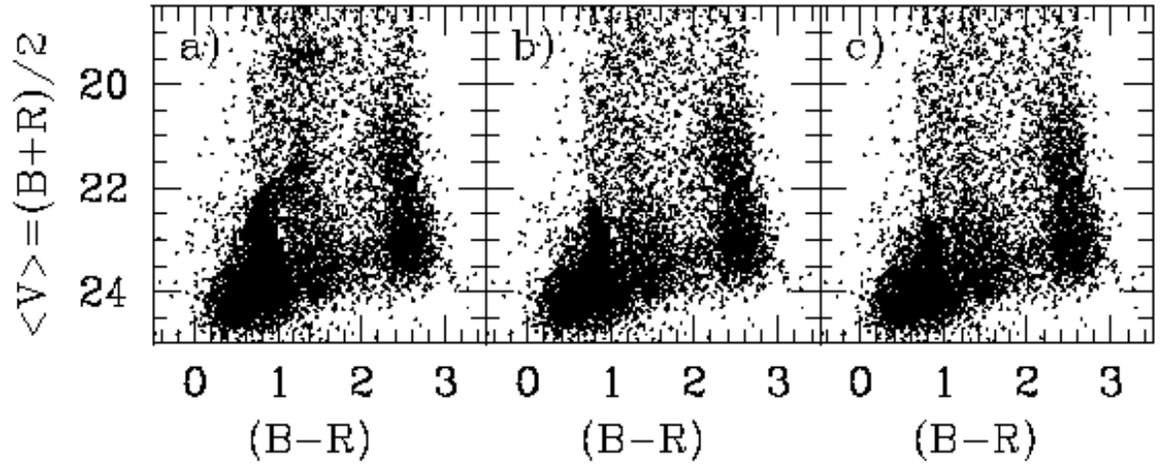}
\caption{ Model CMDs of the Sgr tidal stream for a 
 $35\arcmin \times 35 \arcmin$ field, a distance to the stream of 45 kpc, and surface brightness of
$\Sigma_{V}\sim$ 28.5 mag arcsec$^{-2}$ ({\it panel a}), $\Sigma_{V}\sim$ 30.0 mag arcsec$^{-2}$ 
({\it panel b}), and $\Sigma_{V}\sim$ 30.5 mag arcsec$^{-2}$
({\it panel c}). This model  was made using our control field to represent the typical Milky 
Way halo population for intermediate galactic latitude ($b=50\arcdeg$) and a decontaminated 
CMD of the Sgr main body (see Sec. 2) to represent the stellar population of the stream.\label{fig1}}
\end{figure}

\begin{figure}
\plotone{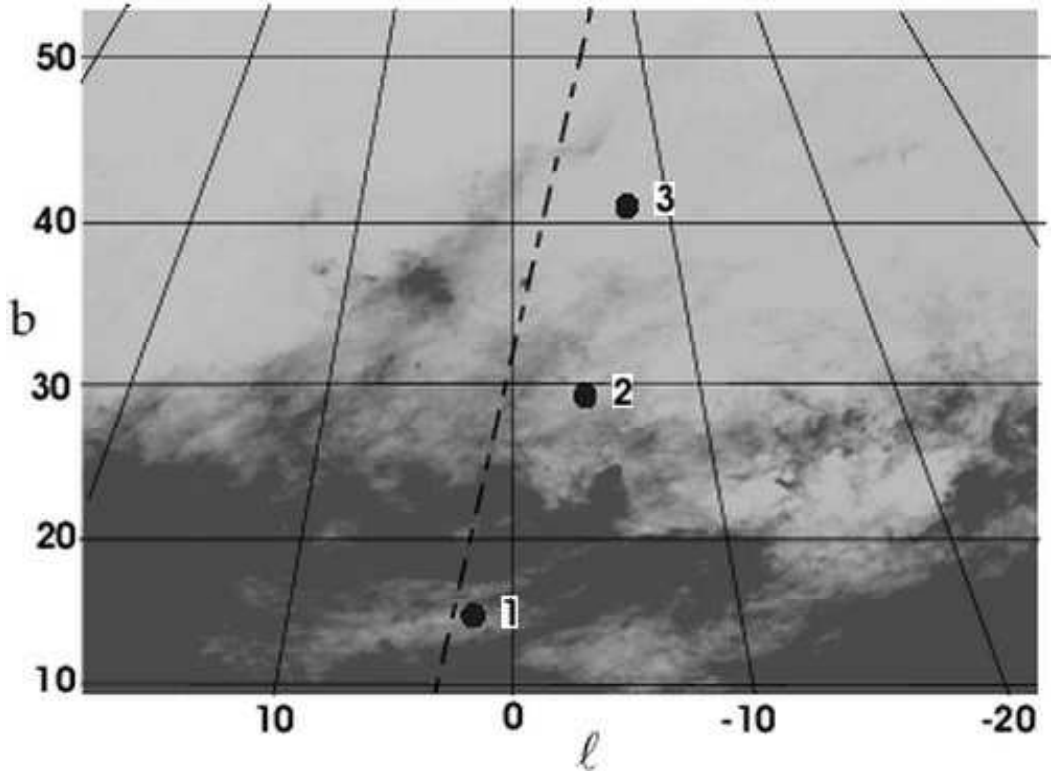}
\caption{The position of the Sgr northern stream predicted by our 
theoretical model in Mart\'\i nez-Delgado et al. (2001a) ({\it dashed line}) 
overplotted to the reddening map of SFD. The positions of the target fields 
discussed in the present paper (see Table 2) are marked with {\it filled 
circles}: SGR2+13 (label 1); SGR357+29 (label 2); and 
SGR352+42 (label 3). White corresponds to low reddening.}
\end{figure}

\begin{figure}
\plotone{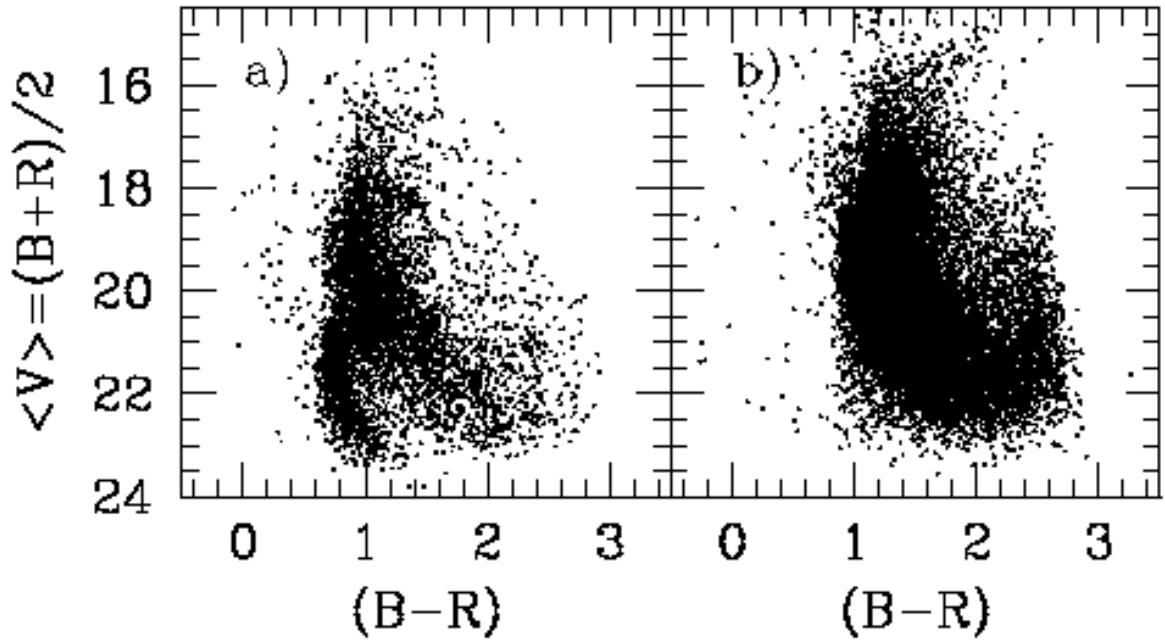}
\caption{CMD of a field situated in the main body of Sgr
({\it a\/}) and of Field 1 ({\it b\/}). Although both diagrams show a similar foreground 
contamination pattern, no signature of 
the Sgr main-sequence turnoff is observed in the CMD of our target field. }
\end{figure}

\begin{figure}
\plotone{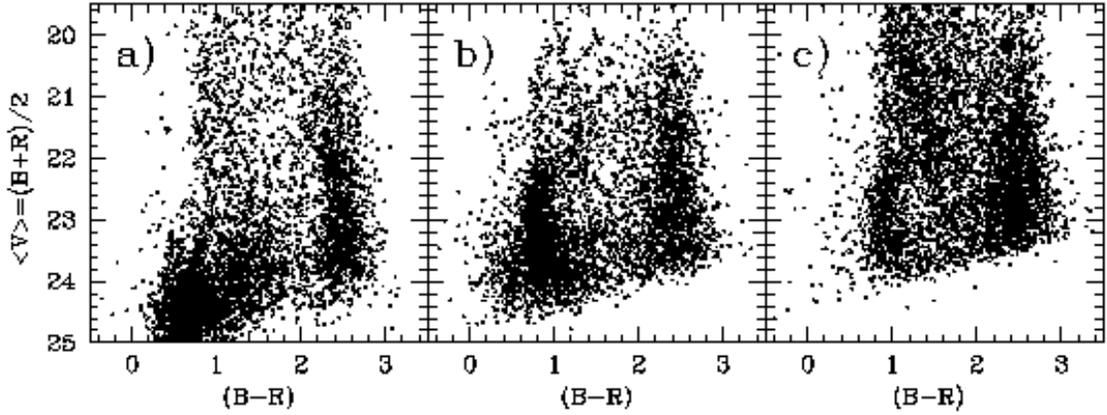}
\caption{CMDs of the control field ({\it panel a\/}), SGR352+42
({\it panel b\/}, labelled as Field 3 in Fig. 2), and SGR357+29 ({\it panel c\/},
labelled as Field 2 in Fig. 2)
(see Table 2). Panel {\it a\/} provides the distribution of the foreground Milky Way 
stars. The overdense strip at $(B-R)\simeq 0.8$, $22.5\leq$''$V$''$\leq 24$ in the CMD of
panel {\it b\/} is interpreted as produced by the Sgr northern
stream. A similar feature is also visible at similar position in the CMD of Field 2 (Panel 
{\it c\/}), situated 
$\sim$$10\arcdeg$ SE from Field 3. The total area covered by  each field is  
$35\arcmin \times 35 \arcmin$.}
\end{figure}

\begin{figure}
\plotone{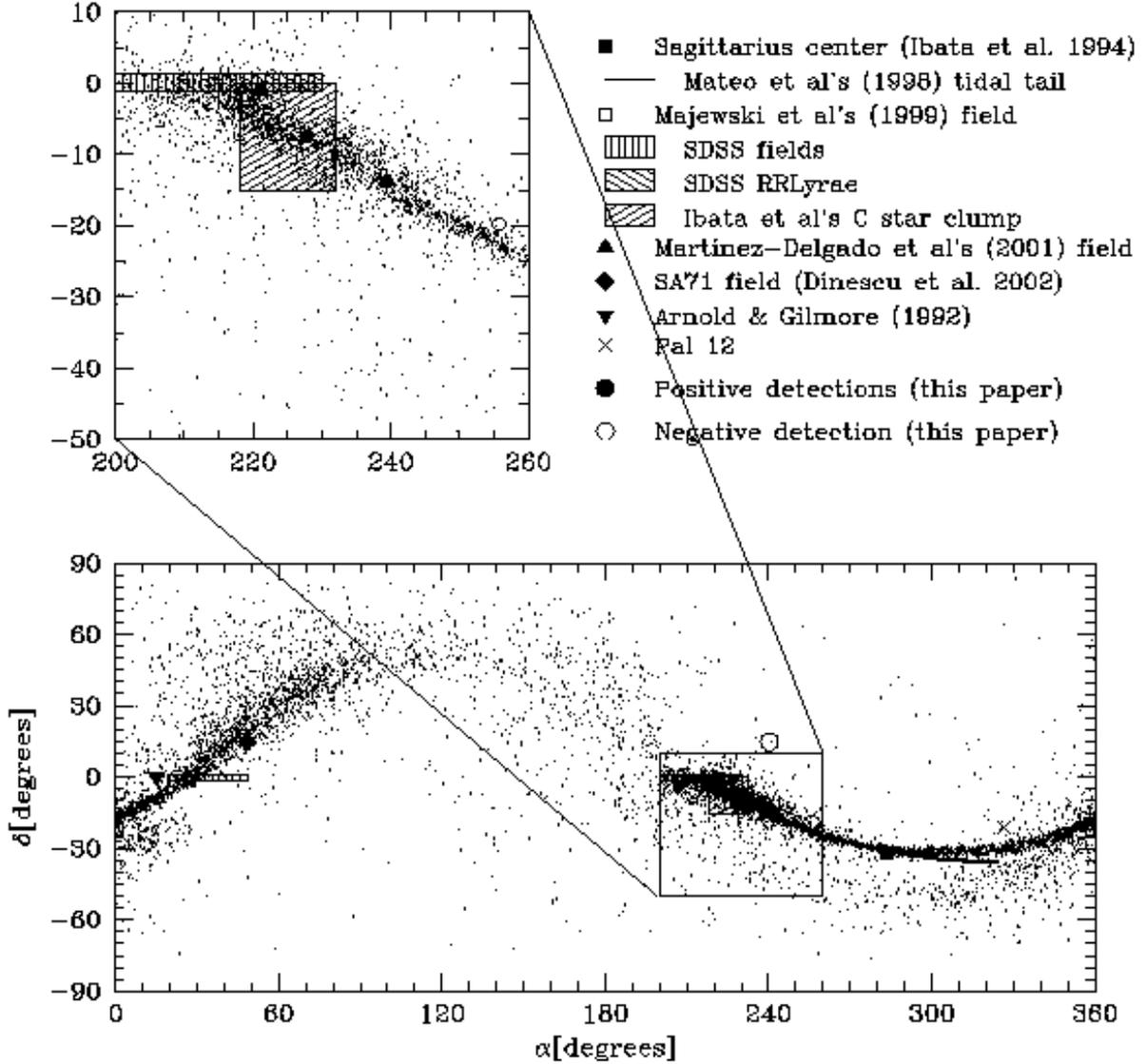}
\caption{({\it lower panel}) Equatorial coordinates ($\alpha$, $\delta$) of our Sgr model 
({\it black dots}) together with the available observational data for the Sgr dwarf galaxy and 
its stream (a detailed explanation is given
in  Sec. 4). The upper panel represents a detail of the apocenter region.}
\end{figure}

\begin{figure}
\plotone{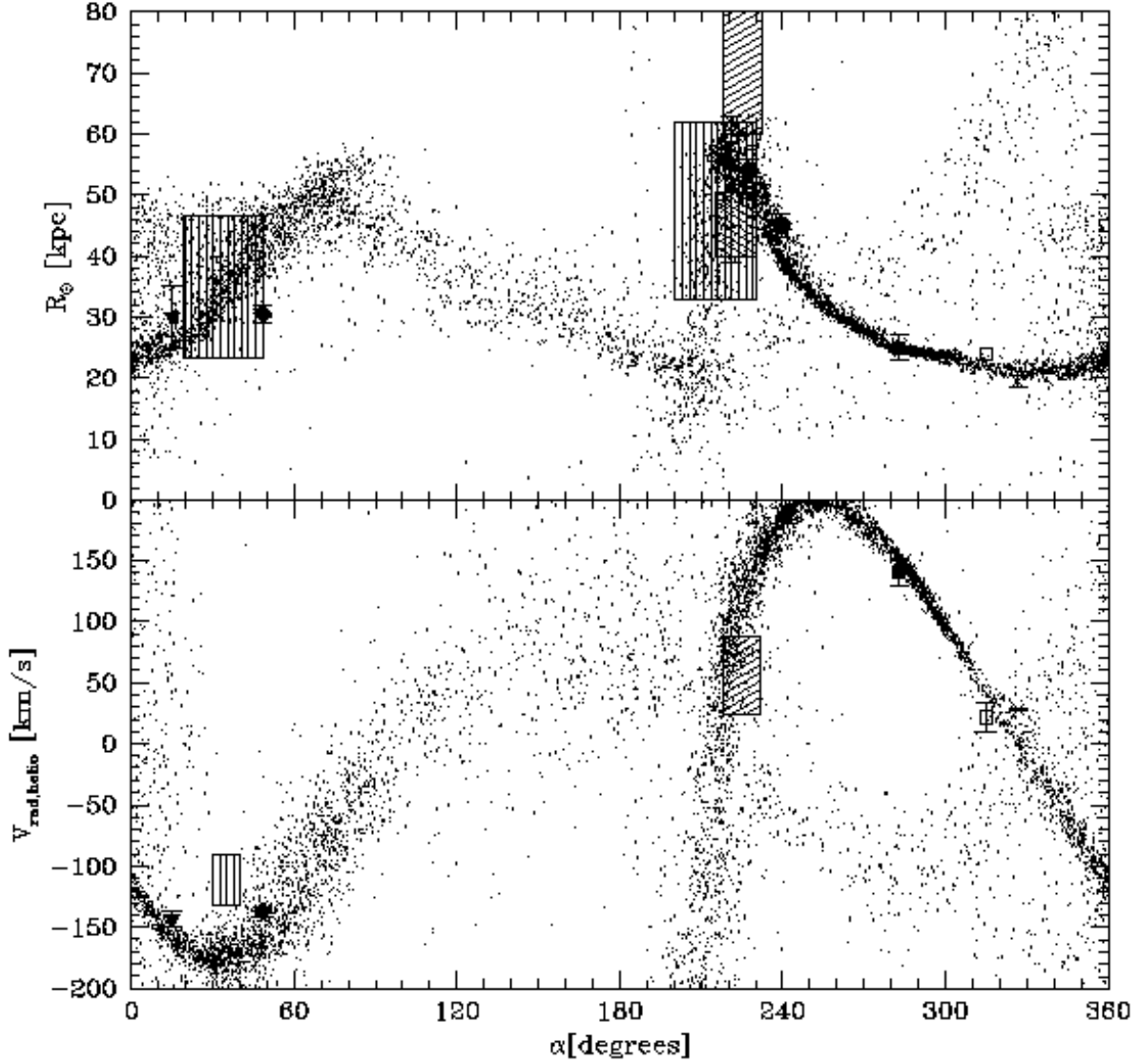}
\caption{Heliocentric distances vs. $\alpha$ of our Sgr model ({\it black dots}) 
together with the observational data related with the Sgr dwarf galaxy ({\it upper panel}). 
Symbols and shaded areas for the observational data of the Sgr tidal stream are the same 
as used in Fig. 5. In the lower panel, the heliocentric radial velocities vs. $\alpha$ for 
our Sgr model and the observational data are plotted. Symbols are the same as in Fig. 5.}
\end{figure}

\begin{figure}
\plotone{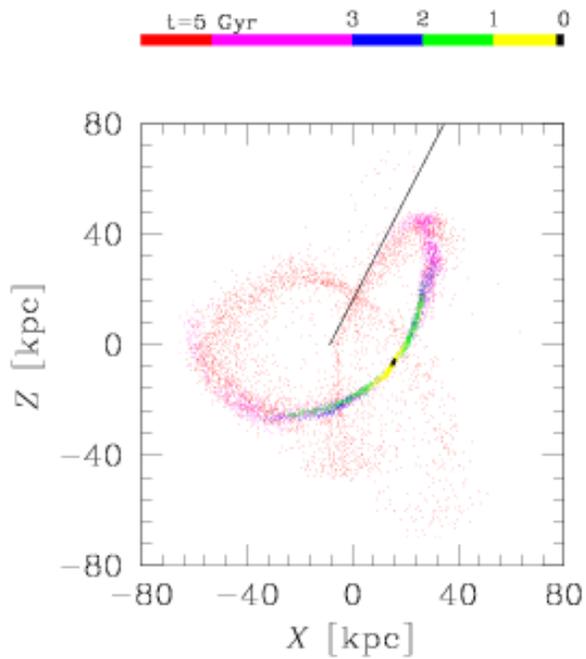}
\caption{XZ-projection of the Sgr stream model  with respect to the
galactic center. The Sun's coordinates are $(X,Y,Z)_{\odot}=(-8.5,0.0,0.0)$ kpc
and Sgr center is placed at $(X,Y,Z)_{Sgr}=(16,2,-5.9)$ kpc. The black particles are still bound to 
the Sgr galaxy; the yellow particles became unbound during the last Gyr;
the green particles became unbound between 1.0 and 2.0 Gyr ago; the blue particles,
between 2.0 and 3.0 Gyr ago; the purple particles, between 3.0 and 5.0 Gyr ago; and the red 
particles, more than 5.0 Gyr ago.}
\end{figure}

\begin{figure}
\plotone{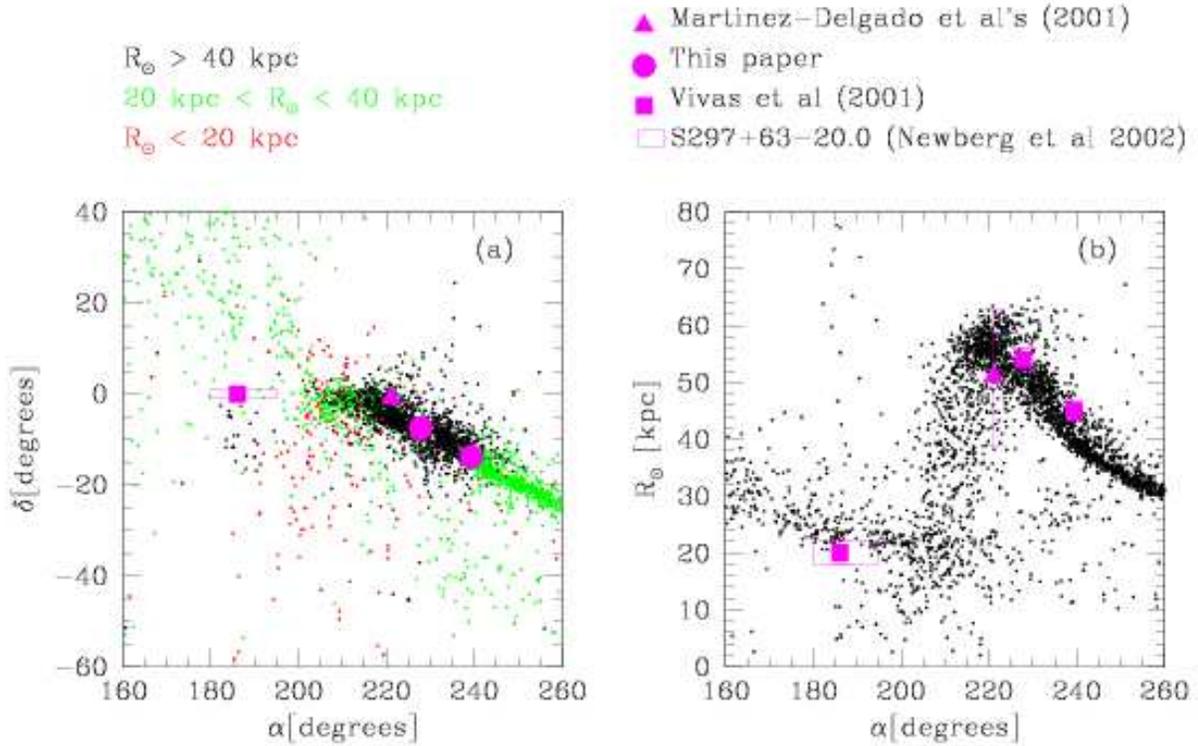}
\caption{(a) Equatorial coordinates $(\alpha, \delta)$ of the Sgr northern stream. The numerical
model of Sgr is plotted as function of the heliocentric distance: the red particles are
closer than 20 kpc; the green particles between 20 and 40 kpc; and the black particles are
further than 40 kpc. In the same panel some observational data are also plotted.
The detail description of these observations is given in Sec. 4.3.
(b) Heliocentric distance vs. $\alpha$ for our Sgr model (black dots) and for some
observational data (described in detail in Sec. 4.3) in the region of the Sgr northern
stream. The crossing of the trailing and leading streams can be observed at $\alpha 
\sim 200-220$.}
\end{figure}

\begin{figure}
\plotone{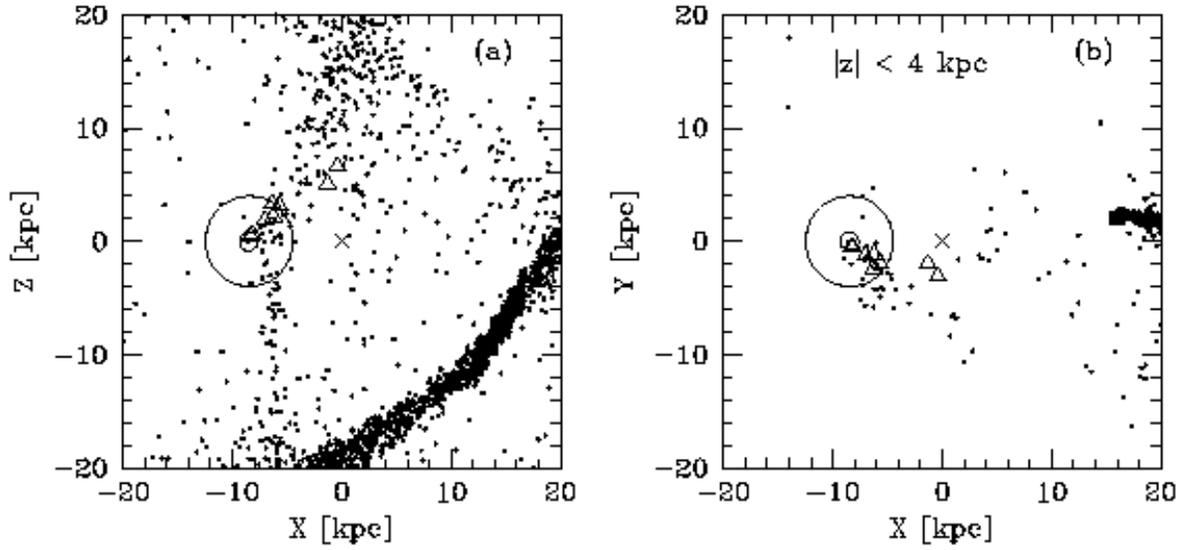}
\caption{(a) XZ-projection in Galactic coordinates of our Sgr model (black dots) in the
solar neighborhood. Sgr is represented with a red square; the Galactic center is the
blue star; the Sun is the open red circle; and the open purple triangles are the
giant stars reported by Kundu et al (2002). The large open green circle marks the limit
of 4 kpc in heliocentric distance.
(b) XY-projection in Galactic coordinates of our Sgr model (black dots), restricted to
the particles with $|Z| < 4$ kpc. Symbols represent the same observational data 
described in panel (a).}
\end{figure}

\begin{figure}
\plotone{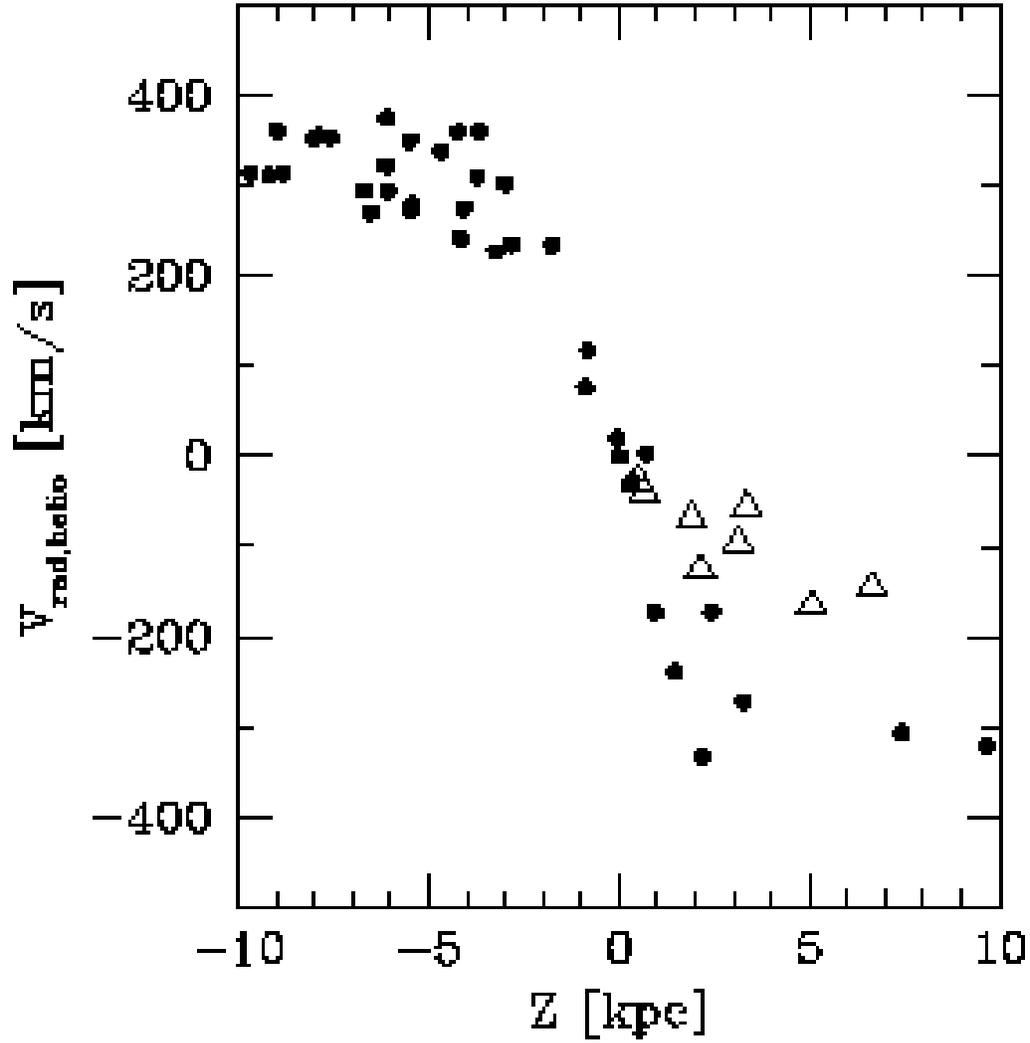}
\caption{Heliocentric radial velocity vs. the vertical distance to the Galactic plane
for our Sgr model (filled circles). The giant stars reported by Kundu et al. (2002) are 
represented by open triangles.}
\end{figure}

\begin{figure}
\plotone{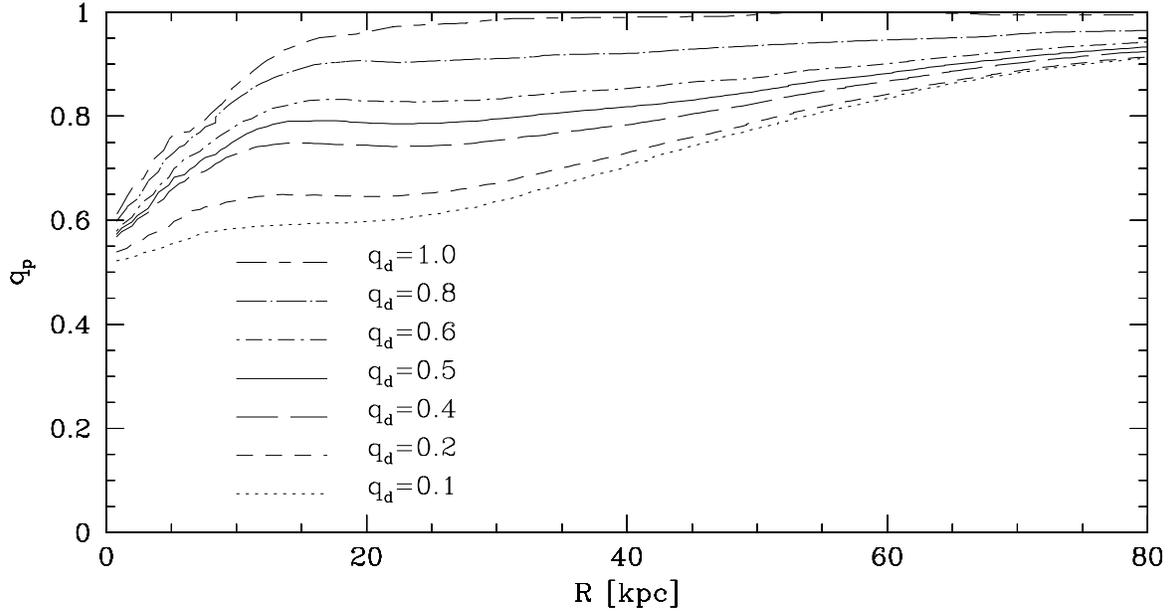}
\caption{Potential flatness, $q_p$,  of the Milky Way for various density flatness, $q_d$, as
a function of distance to the Galactic center.}
\end{figure}

\begin{figure}
\plotone{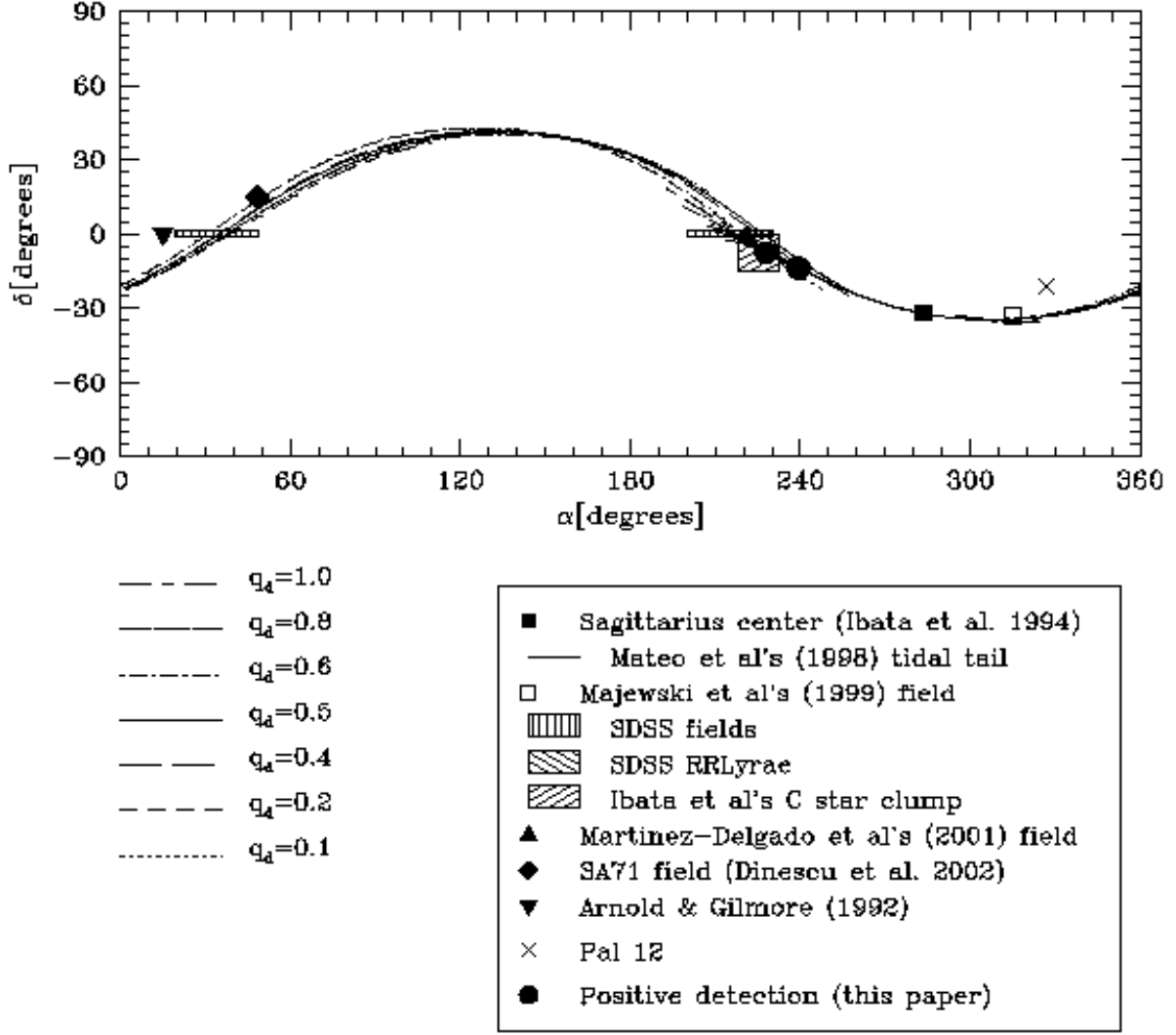}
\caption{Equatorial coordinates ($\alpha$, $\delta$) of the Sgr stream detections 
(symbols are the same as used in Fig. 5) and of the theoretical Sgr orbit for various 
halo flatness, $h=1.0, 0.8, 0.6, 0.5, 0.4, 0.2$ and $0.1$ (lines). }
\end{figure}

\begin{figure}
\plotone{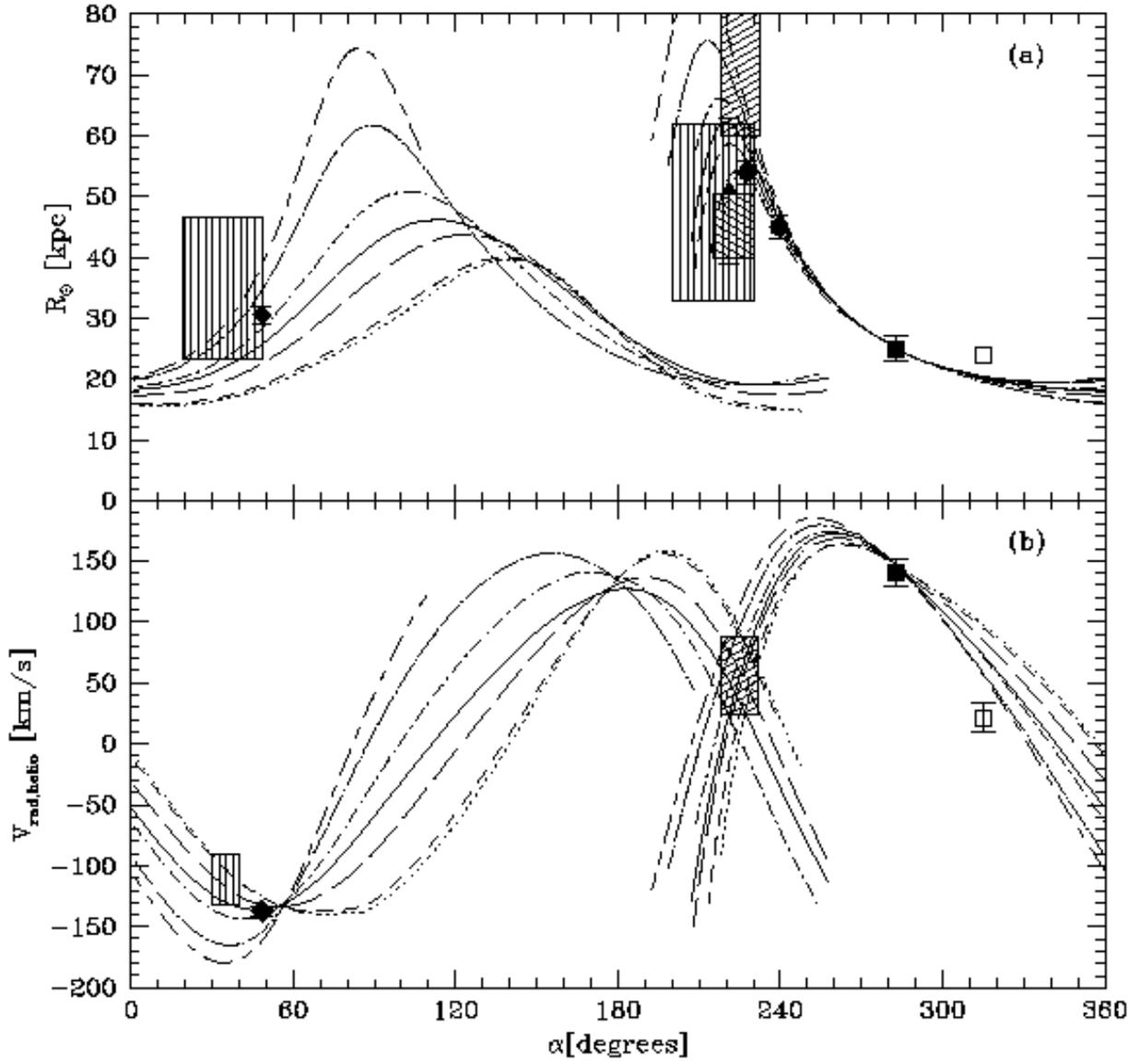}
\caption{(a) Heliocentric
distances vs. $\alpha$ for the Sgr stream detections and for the Sgr orbit in halos with
various flatness (same symbols as Fig. 11). (b) Radial velocities vs. $\alpha$ for the
Sgr stream detections and the Sgr orbit in halos with various flatness.}
\end{figure}

\begin{figure}
\plotone{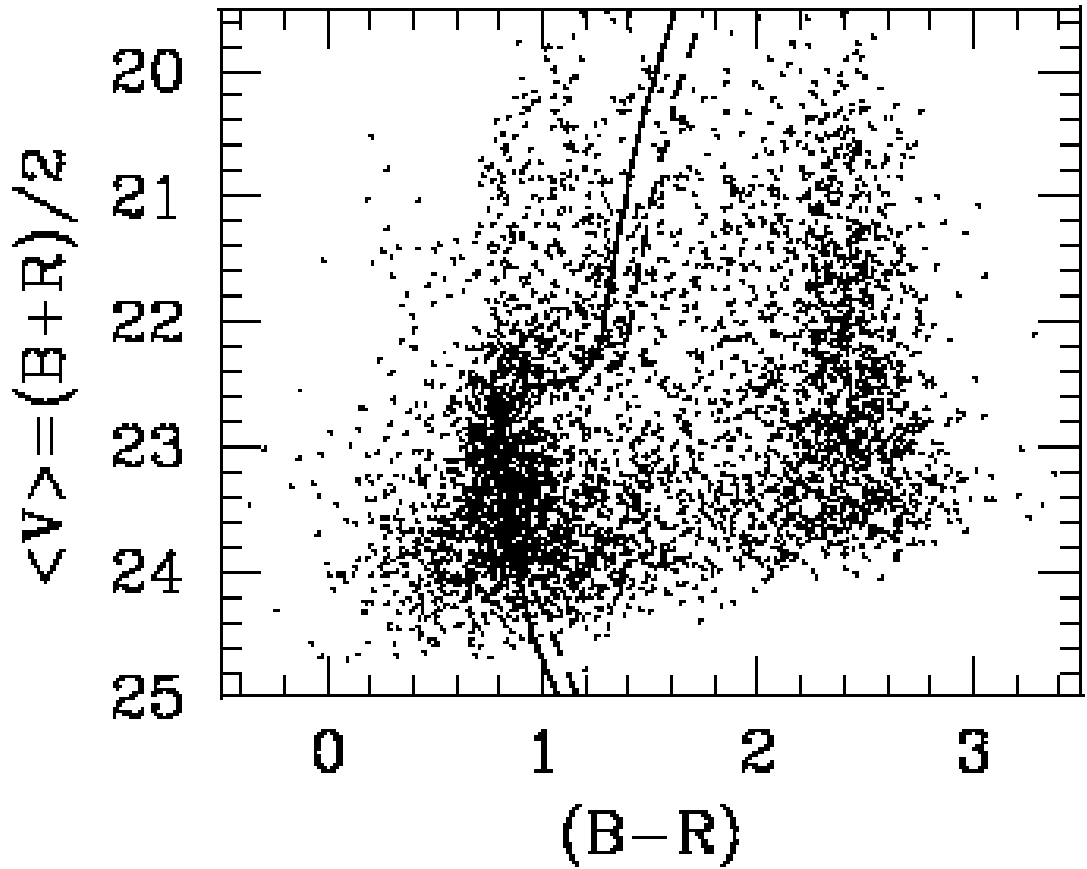}
\caption{CMD of the Sgr northern stream in the SGR352+42 field. 
Isochrones from the Padova library (Bertelli et al. 1994) of 
$Z=0.001$ ([Fe/H] = $-$1.3) and age 11 Gyr ({\it solid line}) and  
$Z=0.004$ ([Fe/H]= $-$0.7) and age 5 Gyr ({\it dashed line}) are overplotted.}
\end{figure}

\begin{figure}
\plotone{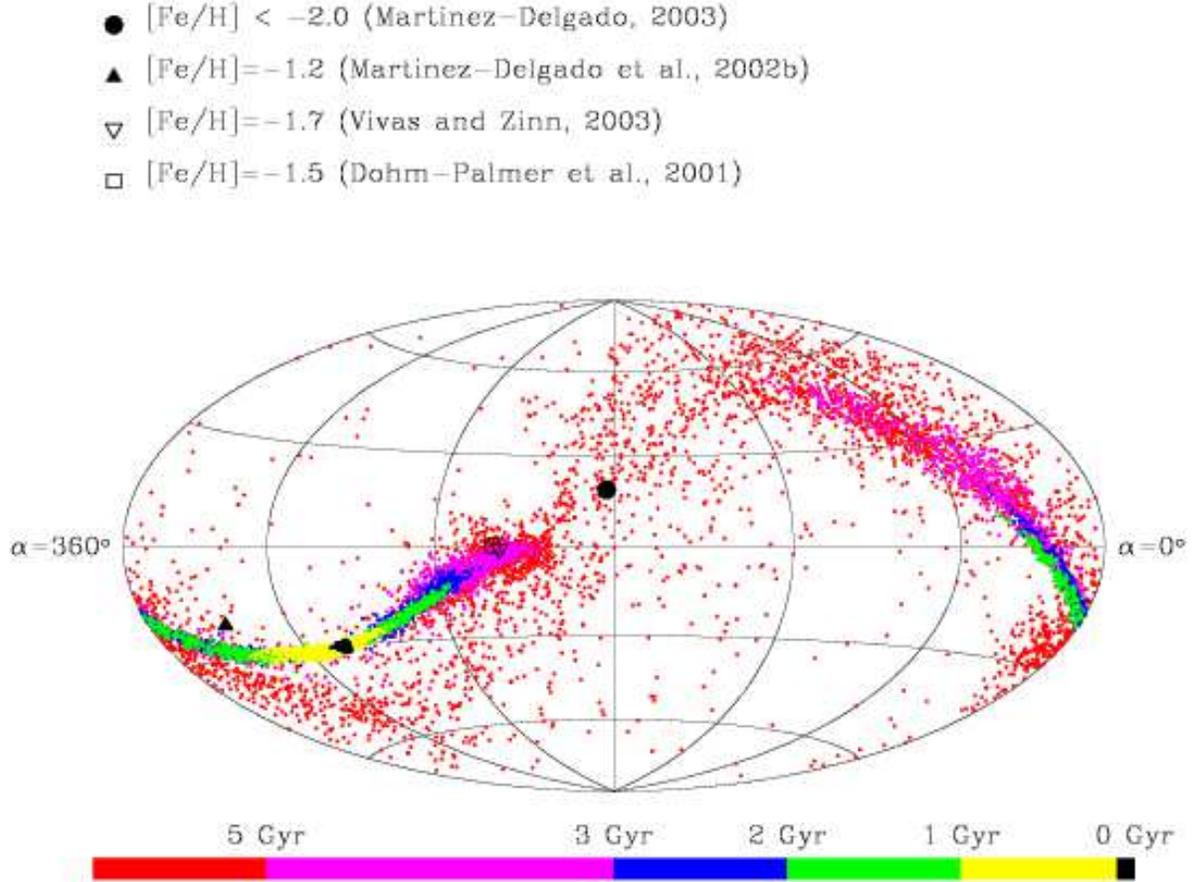}
\caption{Aitoff sky projection in equatorial coordinates of our Sgr model.
Black dots represent particles that are still bound to the Sgr galax; 
the yellow particles became unbound during the last Gyr; the green particles
became unbound between 1.0 and 2.0 Gyr ago; the blue particles, between 2.0
and 3.0 Gyr ago; the purple particles, between 3.0 and 5.0 Gyr ago; and the
red particles, more than 5.0 Gyr ago. Symbols represent observational data
whose metallicity has been measured (see Sec. 6 for a detailed description).}
\end{figure}

\end{document}